\documentclass{aa}
\usepackage[varg]{txfonts}
\usepackage{graphicx,aas_macros,natbib}
\usepackage{subfigure,aas_macros}

\def\stacksymbols #1#2#3#4{\def\theguybelow{#2}
        \def\verticalposition{\lower#3pt}
        \def\spacingwithinsymbol{\baselineskip0pt\lineskip#4pt}
        \mathrel{\mathpalette\intermediary#1}}
\def\intermediary #1#2{\verticalposition\vbox{\spacingwithinsymbol
        \everycr={}\tabskip0pt
        \halign{$\mathsurround0pt#1\hfil##\hfil$\crcr#2\crcr
                \theguybelow\crcr}}}
\def\lta{\stacksymbols{<}{\sim}{2.5}{.2}}
\def\gta{\stacksymbols{>}{\sim}{2.5}{.2}}

\begin{document}

\title{The relation between gas density and velocity power spectra in galaxy clusters: 
high-resolution hydrodynamic simulations \\
and the role of conduction}

\author{M. Gaspari$^1$\thanks{E-mail: mgaspari@mpa-garching.mpg.de} \and E. Churazov$^{1,2}$ \and D. Nagai$^{3,4}$ \and E. T. Lau$^{3,4}$ \and I. Zhuravleva$^{5,6}$} 
\institute{$^1$ Max Planck Institute for Astrophysics, Karl-Schwarzschild-Strasse 1, 85741 Garching, Germany\\
$^2$ Space Research Institute (IKI), Profsoyuznaya 84/32, Moscow 117997, Russia\\
$^3$ Department of Physics, Yale University, New Haven, CT 06520, USA\\
$^4$ Yale Center for Astronomy and Astrophysics, Yale University, New Haven, CT 06520, USA\\
$^5$ Kavli Institute for Particle Astrophysics and Cosmology, Stanford University, 452 Lomita Mall, Stanford, CA 94305-4085, USA\\
$^6$ Department of Physics, Stanford University, 382 Via Pueblo Mall, Stanford, CA 94305-4060, USA
}

%\date{}
%\date{Accepted. Received; in original form}
%\label{firstpage}

\abstract{
Exploring the power spectrum of fluctuations and velocities in the intracluster medium (ICM) can help us to probe the gas physics of galaxy clusters. Using high-resolution 3D plasma simulations, we study the statistics of the velocity field and its intimate relation with the ICM thermodynamic perturbations. The normalization of the ICM spectrum (related to density, entropy, or pressure fluctuations) is linearly tied to the level of large-scale motions, which excite both gravity and sound waves due to stratification. For low 3D Mach number $M\sim0.25$, gravity waves mainly drive {\it entropy} perturbations, traced by preferentially tangential turbulence. For $M> 0.5$, sound waves start to significantly contribute, passing the leading role to compressive {\it pressure} fluctuations, associated with isotropic (or slightly radial) turbulence. Density and temperature fluctuations are then characterized by the dominant process: isobaric (low $M$), adiabatic (high $M$), or isothermal (strong conduction). Most clusters reside in the intermediate regime, showing a mixture of gravity and sound waves, hence drifting towards isotropic velocities. Remarkably, regardless of the regime, the variance of density perturbations is comparable to the 1D Mach number, $M_{\rm 1D}\sim\delta\rho/\rho$. This linear relation allows to easily convert between gas motions and ICM perturbations ($\delta\rho/\rho<1$), which can be exploited by the available {\it Chandra}, {\it XMM} data and by the forthcoming {\it Astro-H} mission. At intermediate and small scales (10$\,$-$\,$100 kpc), the turbulent velocities develop a tight Kolmogorov cascade. The thermodynamic perturbations (which can be in general described by log-normal distributions) act as effective tracers of the velocity field, broadly consistent with the Kolmogorov-Obukhov-Corrsin advection theory. The cluster radial gradients and compressive features induce a flattening in the cascade of the perturbations. Thermal conduction on the other hand acts to damp the thermodynamic fluctuations, washing out the filamentary structures and steepening the spectrum, while leaving unaltered the velocity cascade. The ratio of the velocity and density spectrum thus inverts the downtrend shown by the non-diffusive models, widening up to $\sim$5. This new key diagnostic can robustly probe the presence of conductivity in the ICM. We produce X-ray images of the velocity field, showing how future missions (e.g.~{\it Astro-H}, {\it Athena}) can detect velocity dispersions of a few 100 km s$^{-1}$ ($M>0.1$ in massive clusters), allowing to calibrate the linear relation and to constrain relative perturbations down to just a few per cent.}


\authorrunning{Gaspari et al.}
\titlerunning{The ICM power spectrum -- gas perturbations trace velocities while decouple via conduction} 

\maketitle

\section{Introduction}\label{s:intro}

The power spectrum of perturbations and velocity in a given fluid has historically represented one of the crucial 
tools to understand and constrain the dominant astrophysical processes.
In the cosmology field, the temperature fluctuations in the cosmic microwave background have allowed to put 
precise constraints on the geometry and composition of the universe (e.g.~through the acoustic spectral peaks; \citealt{Planck:2013_spectrum}). 
Closer to our case, the observed electron density perturbations in the interstellar plasma (ISM) have revealed a highly turbulent medium, showing a Kolmogorov power-law spectrum spanning more than 10 decades (\citealt{Armstrong:1981,Armstrong:1995} and references therein). The observed power spectrum of the solar wind density, also consistent with the famous $-5/3$ slope, has further proven that turbulent processes are a key component shaping the dynamics of astrophysical plasmas (e.g.~\citealt{Woo:1979,Marsch:1990}).
In a similar way, the wealth of information contained in the power spectrum
extracted from the hot plasma filling galaxy clusters
can help us to significantly advance our knowledge of the ICM astrophysics.

In the context of galaxy clusters,
\citeauthor{Gaspari:2013_coma} (2013; hereafter GC13)
have shown for the first time that the power spectrum of the ICM density fluctuations linearly rises with the level of 
turbulent motions. Diffusive processes, as thermal conduction, instead fight to damp the cascade of perturbations. 
Many questions still remain to be tackled. 
In this work, we focus on the statistics and features of the {\it velocity field} in the stratified intracluster medium, such as the power spectrum, the real-space and projected maps, and in particular its intimate relation with the thermodynamic perturbations.
The ICM power spectrum can be viewed in various forms (e.g.~\citealt{Schuecker:2004,Churazov:2012}), through the lenses of gas velocities ($\delta v/c_{\rm s}$), density ($\delta\rho/\rho$), entropy ($\delta K/K$), or pressure ($\delta P/P$) fluctuations, thus offering multiple joint constraints. Each physical process leaves marked imprints behind.
The normalization of the perturbation spectrum is tied to the combined action of gravity and sound waves excited by large-scale (100s kpc) turbulence. At intermediate scales, the thermodynamic perturbations act as `tracers' of the eddy inertial cascade (in line with the classic advection theory by \citealt{Obukhov:1949} and \citealt{Corrsin:1951}), while rising diffusivity conspire to decouple the tight relation.
Faster gas motions alter the thermodynamic mode (from isobaric to adiabatic), changing the interplay between different fluctuations.

The linear relation between the density variance and the turbulent Mach number $M$ has been also observed in simulations of supersonic isothermal turbulence in homogeneous and periodic boxes (e.g.~\citealt{Padoan:1997,Konstandin:2012} and references therein), in connection with ISM studies. However, this relation purely arises from the high compression imparted by supersonic turbulence, creating shocks and sharp peaks (e.g.~\citealt{Kim:2005}). In the subsonic regime, the compressibility drastically diminishes, and density perturbations fade as $M^2$ (\citealt{Kowal:2007}).
In the case of stratified galaxy clusters, our novel linear relation is instead tied to the radial gradients of entropy and pressure (\S\ref{s:disc}), already developing in the subsonic regime, i.e.~the realistic state of ICM turbulence ($M\sim0.2-0.7$; e.g.~\citealt{Norman:1999,Lau:2009, Vazza:2009}). 
 
The velocity statistics of the diffuse medium is notoriously difficult to assess through observations. On the contrary, X-ray surface brightness images can robustly constrain the ICM density. 
Being able to convert between the spectra (or even just the normalization) of perturbations and velocities, is a powerful tool, which can be exploited by theoretical studies and by the large amount of available {\it Chandra} and {\it XXM} data.
For instance, the quick estimate of the ICM turbulent velocities allows to study the level of hydrostatic equilibrium in the hot halo (e.g.~\citealt{Vikhlinin:2006}), the transport and dilution of metals (e.g.~\citealt{Rebusco:2005}), the deposition of energy imparted by the active galactic nucleus (AGN) outflows (e.g.~\citealt{Churazov:2004,Gaspari:2011b,Gaspari:2012b}), the evolution of filaments and bubbles (e.g.~\citealt{Scannapieco:2008}), or the reacceleration of cosmic rays (e.g.~\citealt{Brunetti:2007}). 
On the other hand, being able to quickly assess the conductive state of the plasma allows to constrain the survival of the cold/warm gas, which is crucial for star formation (e.g.~\citealt{McDonald:2009}) and black hole accretion (e.g.~\citealt{Gaspari:2013_cca}), to study the quenching of cooling flows (e.g.~\citealt{Kim:2003}) and the evolution of cosmic structures (e.g.~\citealt{Dolag:2004}).

We could soon take advantage of the inverse process, albeit more expensive.
The upcoming {\it Astro-H} mission (\citealt{Takahashi:2010}) and the future {\it Athena} (\citealt{Nandra:2013}) will provide unprecedented detections of the turbulent velocity dispersion in the ICM (via line broadening), as well as bulk motions (via line shift; e.g.~\citealt{Inogamov:2003,Zhuravleva:2012,Nagai:2013,Tamura:2014}), down to Mach numbers $\sim 0.1$ for massive clusters (see \S\ref{s:maps}). 
Reliable constraints on the gas motions allow to accurately calibrate the above relation, and to assess the level of density fluctuations if the imaging is poor. 
For instance, apparently `relaxed' systems may host $\gta10$ per cent density fluctuations, which may significantly alter the formation or regeneration of cool cores in clusters, as well as biasing the estimate of radial profiles, to name a few interesting applications. The same spectral analysis can be extended to the gaseous halos of massive galaxies and groups.

The physics of the intracluster medium is a strongly debated topic. Turbulence has been mainly studied
by means of cosmological and isolated simulations (e.g.~\citealt{Norman:1999, Dolag:2005, Kim:2005, Nagai:2007, Lau:2009, Vazza:2009, Valdarnini:2011, Borgani:2011, Miniati:2014,Schmidt:2014,Shi:2014}). Similarly, diffusion processes as conduction have been mainly investigated via theoretical studies (\citealt{Chandran:1998,Narayan:2001,Zakamska:2003,Ruszkowski:2010,Ruszkowski:2011,Voigt:2004,Roediger:2013_visc,Smith:2013,ZuHone:2013}). Observations have instead hard time in resolving and constraining such processes through local features, granting in the last decade only a few estimates (\citealt{Ettori:2000,Markevitch:2007,Forman:2007}; Eckert et al. 2014, in prep.). However, we are now able to retrieve the statistics of density/pressure fluctuations in the ICM (\citealt{Schuecker:2004,Churazov:2012,Sanders:2012}), allowing to probe the gas physics without the need to resolve local structures.

In \citet{Churazov:2012}, we outlined as possible effects contributing to the density fluctuations: turbulence (via the Bernoulli term $\propto M^2$ or via sound waves), entropy variations (due to mergers or turbulence), perturbations of the gravitational potential, metallicity variations, and AGN bubbles. Using 3D high-resolution plasma simulations, we focus in this work on the role of turbulence and thermal diffusivity, and the driven thermodynamic perturbations, including entropy and pressure variations. Controlled experiments allow us to discriminate the exact contribution of each included physics.
In a companion paper (\citealt{Zhuravleva:2014}, hereafter Z14), we analyze the density perturbations in cosmological AMR simulations, focusing on the role of gravity waves. At the price of lower resolution, we are thus able to include the turbulence driving led by mergers (\S\ref{s:imp}).

This work is structured as follows. 
In \S\ref{s:init}, we review the main physical and numerical ingredients of the simulated models.
In \S\ref{s:spec}, we present the power spectrum of velocities and density fluctuations, focusing on their tight connection and key features (normalization, cascade, damping).
In \S\ref{s:maps}, we analyze the real-space properties of the velocity and perturbation field, showing what X-ray observations are able to detect.
In \S\ref{s:disc}, we throughly discuss the physical interpretation of the power spectrum, as the interplay of $g$-waves and $p$-waves, along with the development of the spectral cascade of all thermodynamic perturbations, in relation to the advection theory of tracers.
In \S\ref{s:conc}, we summarize the results and remark how the ICM power spectrum can be exploited by future observations and theoretical studies, to probe the physics of the diffuse medium with high precision.

\section[]{Physics and numerics} \label{s:init}
The implemented physics and numerics are described in depth in \citeauthor{Gaspari:2013_coma} (2013; section 2), to which the reader is referred for the complete details. Here we summarize the essential features.

The initial conditions for the hot gas are modeled following the latest {\it XMM} observed temperature and density radial profiles of Coma cluster ($\beta-$model with core radius $r_{\rm c} = 272$ kpc and index 0.75). Given the high ICM temperature, $T \sim8.5$ keV, and low electron number density, $n_{\rm e}\sim4\times10^{-3}$, Coma serves as excellent laboratory to study the effects of conduction and turbulence, without being strongly influenced by radiative cooling or AGN feedback. The hot gas is initialized in hydrostatic equilibrium, providing a gravitational potential
appropriate for a massive cluster in the $\Lambda$CDM universe with virial mass $M_{\rm vir}\sim10^{15} M_\odot$ ($r_{\rm 500}\sim1.4$ Mpc, covered by the width of the 3D box).

The conservative hydrodynamics equations are integrated with the Eulerian code FLASH4 (\citealt{Fryxell:2000}), using a third order scheme (piecewise parabolic method) in the framework of the unsplit flux formulation. The ICM plasma has adiabatic index $\gamma=5/3$ and mean atomic weight $\mu\simeq0.62$.
We chose the numerically expensive uniform grid ($512^3$), instead of adaptive cells, in order to remove any substantial, spurious diffusivity due to the refinement/derefinement. The resolution is $\Delta x\,$$\simeq\,$2.6 kpc, roughly on the scale of the (unmagnetized) plasma mean free path, mimicking slightly suppressed Spitzer viscosity. Boundary zones have Dirichlet condition, fixed by the large-scale radial profile; inflow is prohibited.
In addition to hydrodynamics, we add as source terms the turbulence driving, thermal conduction, and electron-ion equilibration. 

Injection of subsonic, solenoidal turbulence is modeled with a spectral forcing scheme that generates a statistically stationary velocity field (GC13, sec.~2.2), based on an Ornstein-Uhlenbeck random process. The amplitudes of the driven acceleration are evolved in Fourier space and then directly converted to physical space. 
Since observations (\citealt{Schuecker:2004, Churazov:2008, DePlaa:2012, Sanders:2013}) and simulations (\citealt{Norman:1999, Lau:2009, Vazza:2009, Vazza:2011, Gaspari:2012b, Schmidt:2014, Shi:2014}) show that the ICM turbulent energies are $\sim3-30$ percent of the thermal energy, we test subsonic Mach numbers in the range $M\equiv\sigma_v/c_{\rm s}\sim0.25-0.75$, where $\sigma_v$ is the 3D\footnote{Volume-weighted or mass-weighted 3D Mach number is very similar, within $\lta 3$ per cent accuracy.} 
velocity dispersion (average sound speed of Coma is $c_{\rm s}\simeq1500$ km s$^{-1}$). The source of turbulence can be various, including cosmological flows/mergers, galaxy motions, and feedback processes.
The former usually dominates, affecting large volumes beyond the core
(e.g. \citealt{Shi:2014}; Z14), and being related to a solenoidal flow (e.g.~\citealt{Miniati:2014}). We thus stir the gas on large scales, with typical injection peak $L\sim600$ kpc (in a few runs $\sim300$ kpc), letting turbulence to naturally cascade.
The turbulence timescale as a function of physical scale $l$ is the eddy turnover time $t_{\rm turb}\simeq (L^{1/3}/\sigma_{v, L})\;l^{2/3}$, using the Kolmogorov scaling $\sigma_v\propto l^{1/3}$. 
Since turbulence is kept subsonic, dissipational heating is subdominant $t_{\rm diss, heat}\sim M^{-2}\,t_{\rm turb}$ (e.g.~\citealt{Ruszkowski:2011}), on timescales of order of the eddy turnover time.
We recall that turbulence acts as an effective diffusivity on entropy with coefficient $D_{\rm turb}\sim \sigma_v\,l$ (e.g.~\citealt{Dennis:2005}).

The conduction of thermal energy, due to the plasma electrons, is driven by a flux 
${\boldsymbol F}_{\rm cond} = -f\,\kappa_{\rm S}\,  \boldsymbol{\nabla} T_{\rm e}$, with conductivity
$\kappa_{\rm S}\simeq 5\times10^{-7}\,T_{\rm e}^{5/2}$  ${\rm erg\, s^{-1}\, K^{-1}\, cm^{-1}}$
(\citealt{Spitzer:1962}).
The related diffusivity and timescale is $D_{\rm cond} = f\,\kappa_{\rm S}/1.5\, n_{\rm e}\, k_{\rm B}$ and $t_{\rm cond}=l^2/D_{\rm cond}$, respectively.
The conductive flux saturates as $F_{\rm sat}\propto n_{\rm e} T^{3/2}_{\rm e}$, whenever the temperature scale height is smaller than the electron mean free path. We use an advanced implicit solver which allows for long, Gyr integration times.
MHD simulations (e.g.~\citealt{Ruszkowski:2010}) show that the outcome of subsonic turbulence is a tangled magnetic field with small kpc coherence length 
(see also constraints in  \citealt{Kim:1990}). On scales larger than the coherence length, the average suppression due to anisotropic conduction and magnetic microinstabilities can be parametrized with the so-called $f$ factor, commonly $f\sim10^{-3}-10^{-1}$ (GC13, sec.~ 2.1.1). 
Using the effective isotropic conductivity has the advantage of modeling any level of suppression affecting the bulk of the ICM.
MHD runs only provide the geometric suppression above the plasma mean free path $\lambda$, in a chaotic atmosphere typically $f\sim1/3$ (\citealt{Ruszkowski:2010}; see also \citealt{Narayan:2001}). However, line wandering and magnetic mirrors, together with plasma microinstabilities, well below $\lambda$ can strongly suppress the transport of heat down to $f\sim10^{-3}$ (\citealt{Rechester:1978, Chandran:1998, Komarov:2014}).
The survival of cold fronts, bubbles, and cold gas (\S\ref{s:intro}), together with our GC13 spectral analysis, point towards strongly suppressed values, $f\sim10^{-3}$.

We integrate both the electron and ion temperature equation, since equilibration times can become considerable in a hot plasma ($t_{\rm ei}\gta50$ Myr). The heat exchange rate is $\propto (T_{\rm e}-T_{\rm i})/t_{\rm ei}$, using Spitzer equilibration time $t_{\rm ei}\propto T_{\rm e}^{3/2}/n_{\rm e}$ (cf.~GC13). The 2T modeling allows to prevent the formation of spurious perturbations due to the unphysical instantaneous transfer of heat.

\section[]{The ICM power spectrum: velocity and $\delta\rho/\rho$} \label{s:spec}

\begin{figure*} 
    \centering
       \subfigure{\includegraphics[scale=0.54]{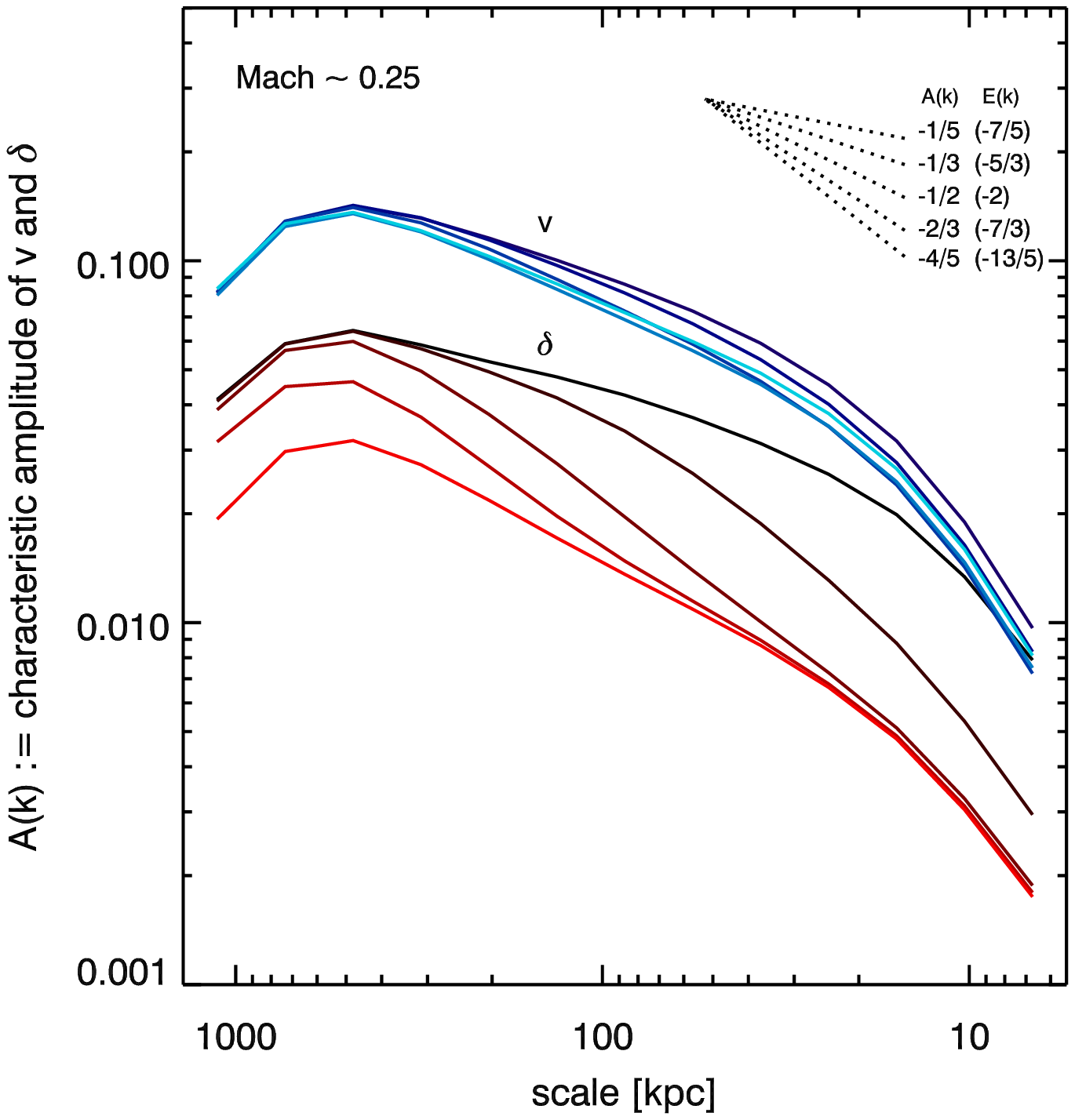}}
       \subfigure{\includegraphics[scale=0.54]{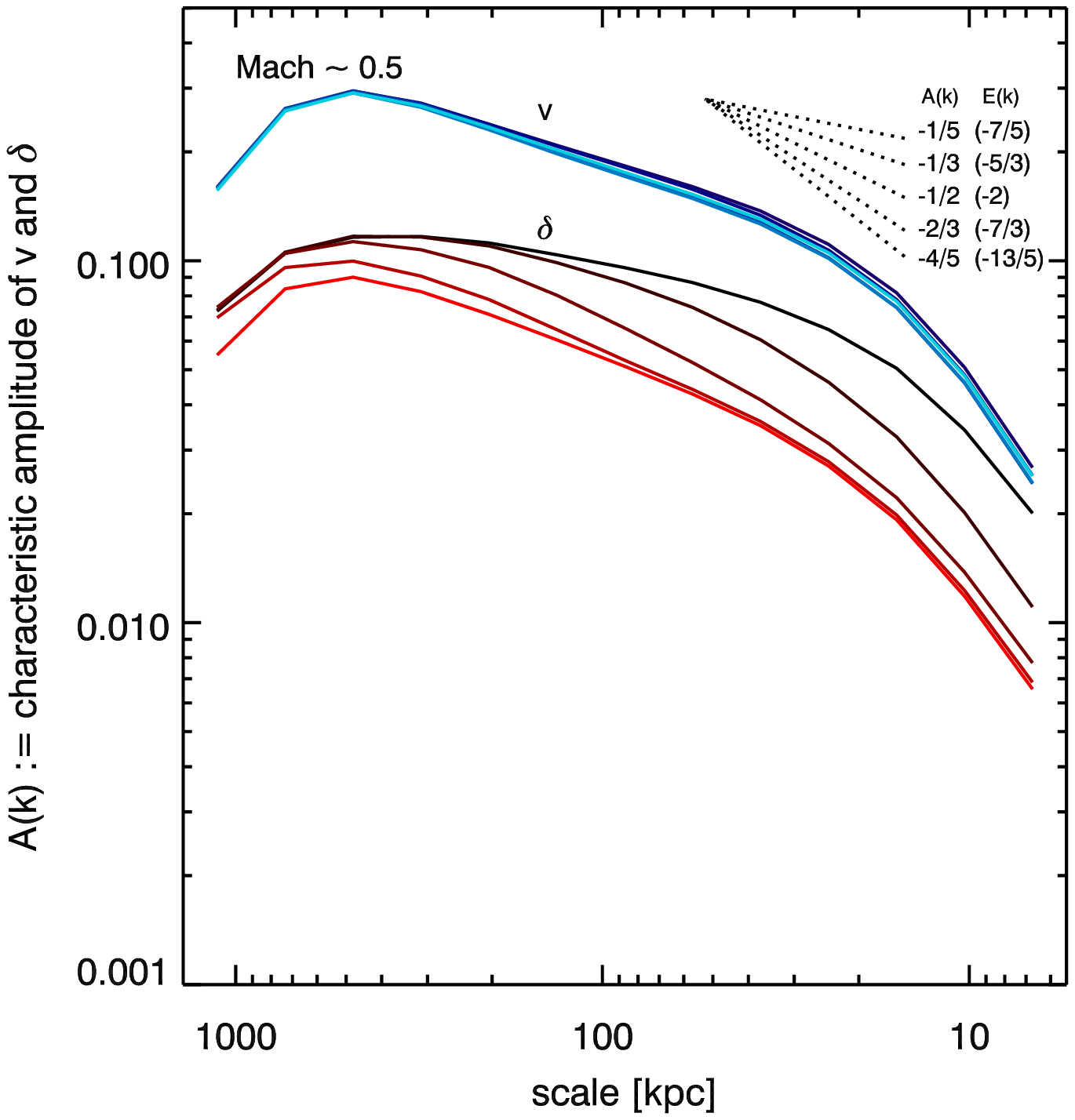}}
       \subfigure{\includegraphics[scale=0.54]{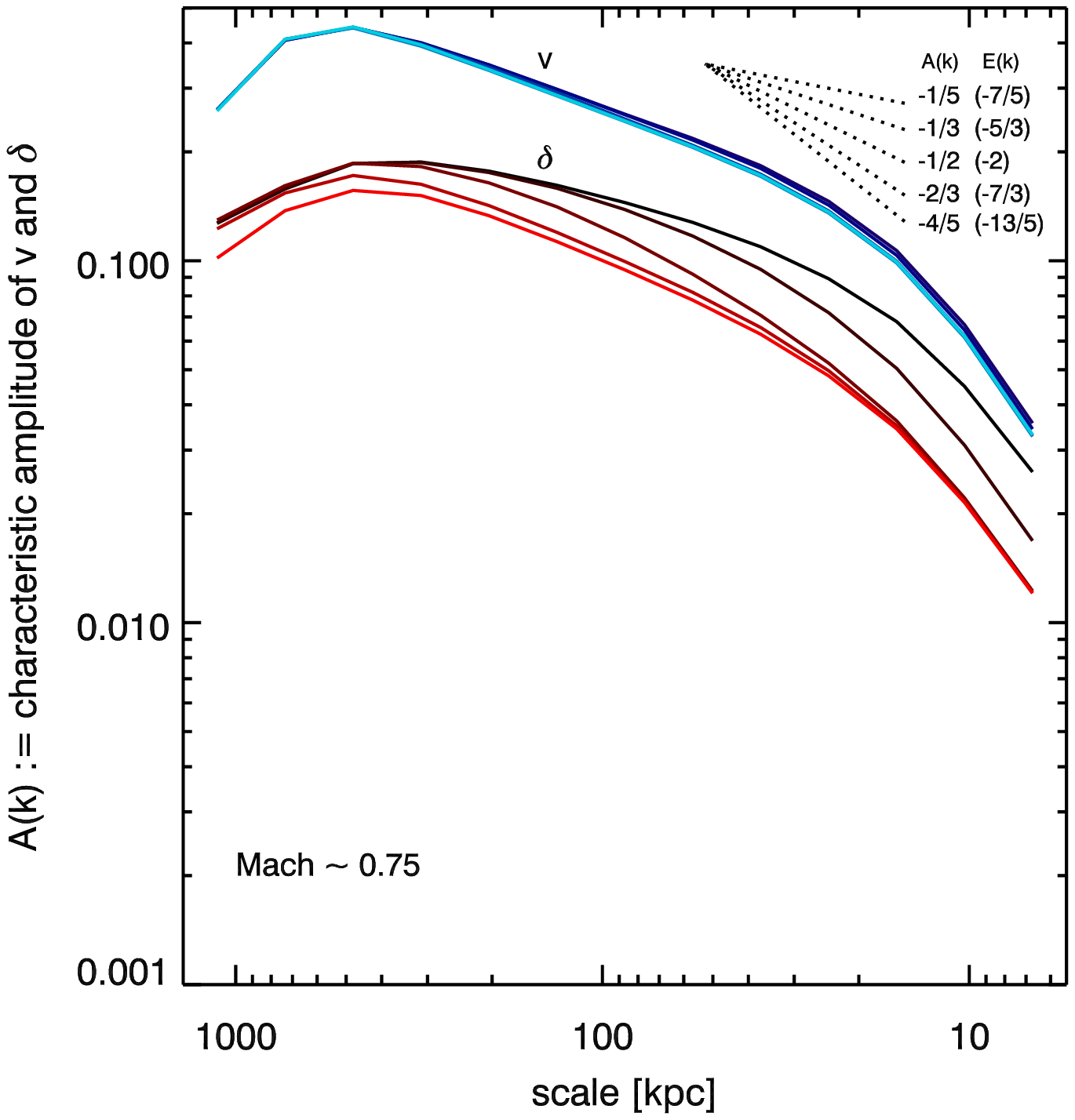}}
       \subfigure{\includegraphics[scale=0.54]{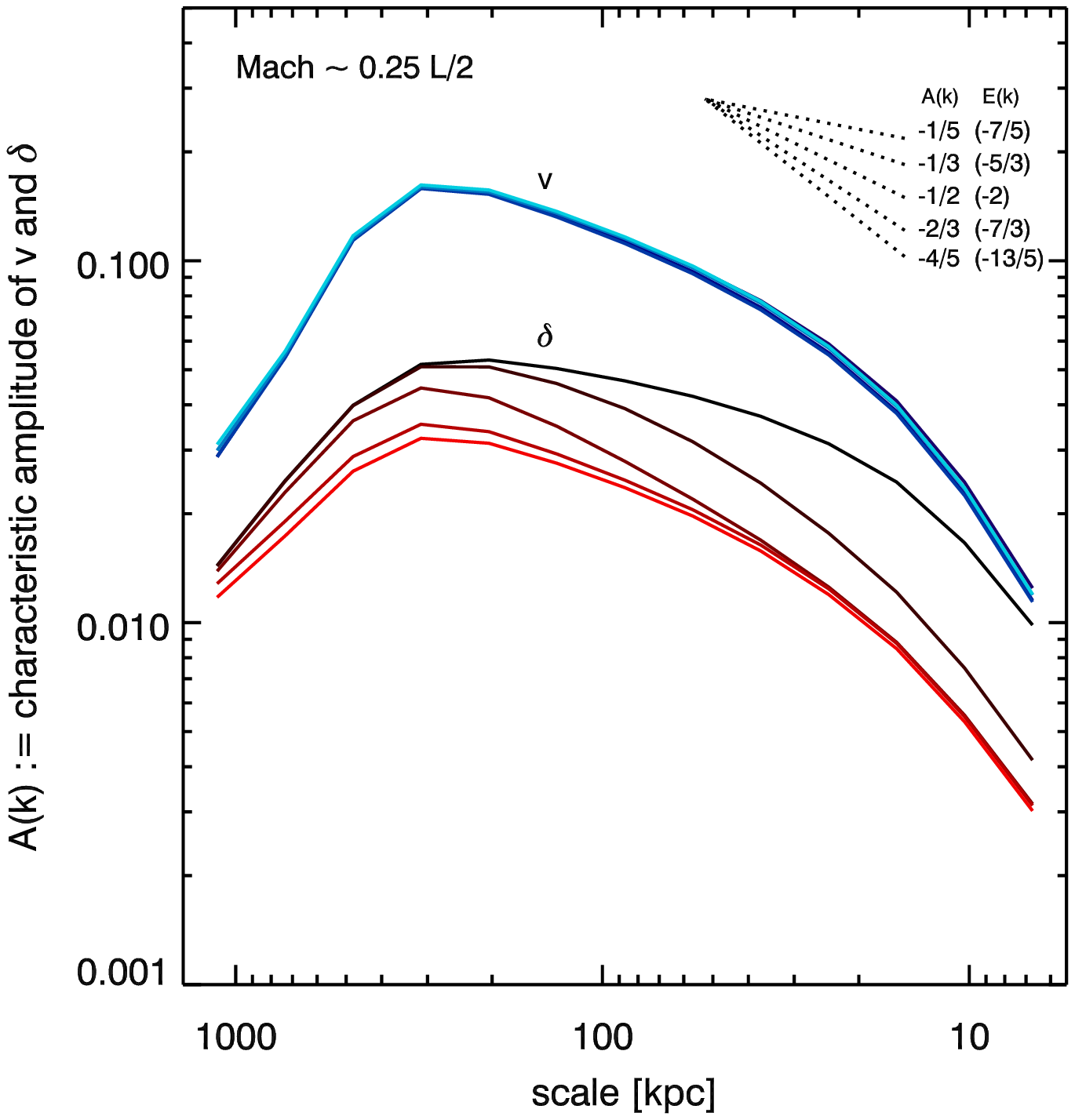}}
       \caption{Characteristic amplitude of $\delta \rho/\rho$ (red) and $v/c_{\rm s}$ (blue), $A(k)=\sqrt{P(k)\,4\pi k^3}$, after reaching statistical steady state ($\sim$$2\ t_{\rm turb}$) with the same level of continuous stirring. From top left: models with weak ($M\sim0.25$), mild ($M\sim0.5$), and strong ($M\sim0.75$) turbulence; the last model (bottom right) has half the reference injection scale ($\sim600/2$ kpc), using $M\sim0.25$. From dark to bright line color, the level of conduction increases by a factor of 10: $f=0\ ({\rm hydro}),\, 10^{-3},\, 10^{-2},\, 10^{-1},\, 1$.
       The evolution is overall self-similar, varying the strength of turbulence or the injection scale.
       Density perturbations are an effective tracer of the velocity field, especially on large scales, with normalization $A_{v_{\rm 1D}}\approx 1.3\, A_\rho $ (at $L\sim600$ kpc). On smaller scales, $\delta\rho/\rho$ displays a cascade shallower than the Kolmogorov slope followed by velocities. Remarkably, conduction strongly damps density perturbations, but leaves unaltered the velocity cascade, thus inverting the $A_v(k)/A_\rho(k)$ ratio (Fig.~\ref{fig:ratio}). 
       }
       \label{fig:Ak}
\end{figure*} 

We now describe the results of the simulated models, focusing on the spectral and real-space properties of the 
turbulent {\it velocity}, in relation with the statistics of gas density perturbations, $\delta \rho/\rho$.
Gas density is indeed the primary astrophysical observable, directly extracted from the X-ray surface brightness.
As thoroughly discussed in \S\ref{s:disc}, the perturbations would be actually more evident through
entropy (for low $M$) or pressure (for high $M$), and then retrieving $\delta \rho/\rho$ via the main thermodynamic mode
(isobaric, isothermal, or adiabatic). Unfortunately, $K$ and $P$ are difficult X-ray observables to constrain.
Nevertheless, although the underlying cause differs, the spectrum of the leading `tracer' is tied to velocities in a very similar manner, granting a fairly universal $M - \delta \rho/\rho$ relation (\S\ref{s:gw}-\ref{s:pw}).  

We first retrieve the characteristic amplitude of total velocity,
normalized to $c_{\rm s}\simeq1500$ km s$^{-1}$.
It is convenient to use the characteristic amplitude, instead of the power spectrum $P(k)$ or energy spectrum $E(k)$,
since its units are the same of the variable in real space.
The amplitude spectrum is defined as
\begin{equation}\label{e:Ak}
A(k) \equiv \sqrt{P(k)\, 4\pi k^3} \equiv \sqrt{E(k)\, k}, 
\end{equation}
where $k=\sqrt{k_x^2+k_y^2+k_z^2}\equiv l^{-1}$ (kpc$^{-1}$).
No major bulk motion is present in our box (average velocity $\sim\,$0); the velocity dispersion is strictly associated with the turbulence driving. 
The relative perturbations instead require to be divided by the underlying background radial profile, e.g.~for density
$\delta \rho/\rho = \rho/\rho_{\rm b} -1$ (GC13, sec.~2.7). 
Except for mild deviations (\S\ref{s:interp}), the turbulence field can be considered isotropic as a first order approximation,
allowing to use the conversion $v_{\rm 1D}\sim v/\sqrt{3}$.
All the power spectra are
computed with the `Mexican Hat' filtering (\citealt{Arevalo:2012}) instead of performing Fourier transforms (GC13, appx.~A for a comparison), which can lead to spurious features due to the box non-periodicity.

\begin{figure} 
    \begin{center}
       \subfigure{\includegraphics*[scale=0.54]{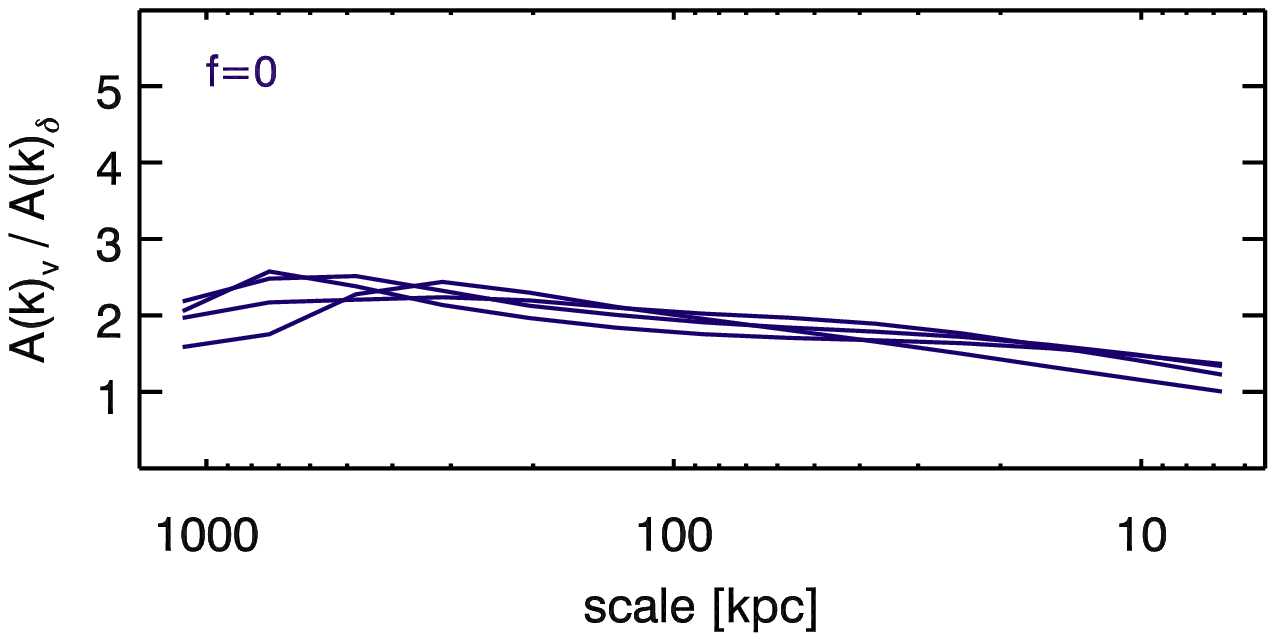}}
       \subfigure{\includegraphics*[scale=0.54]{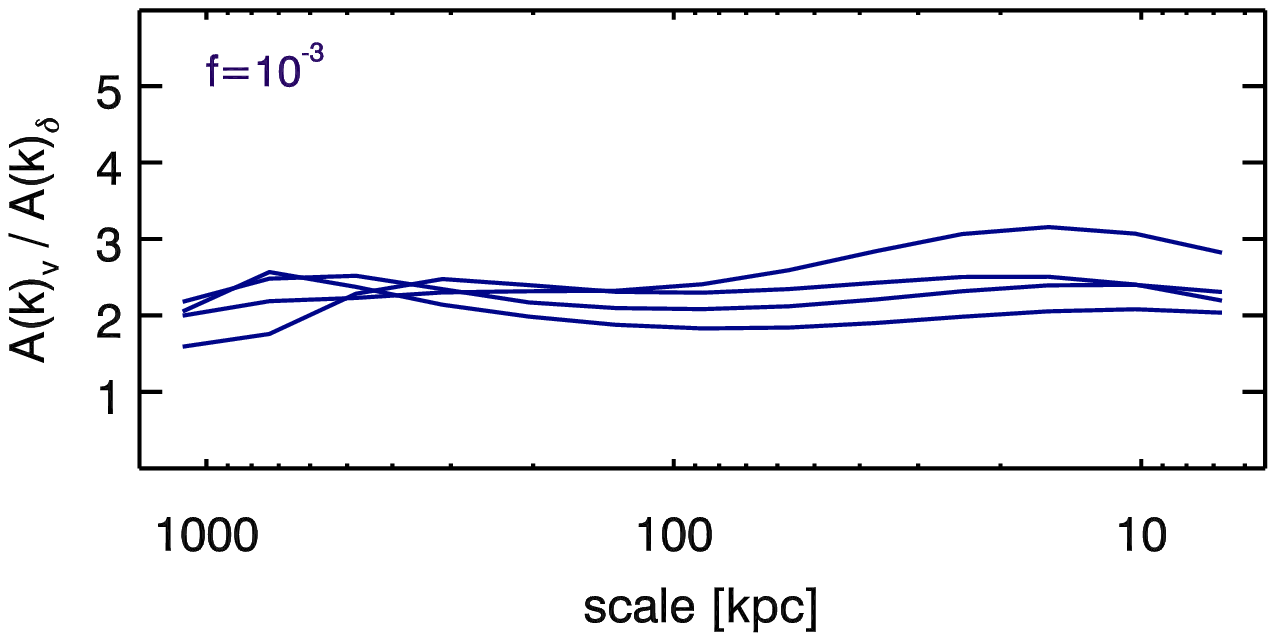}}
       \subfigure{\includegraphics*[scale=0.54]{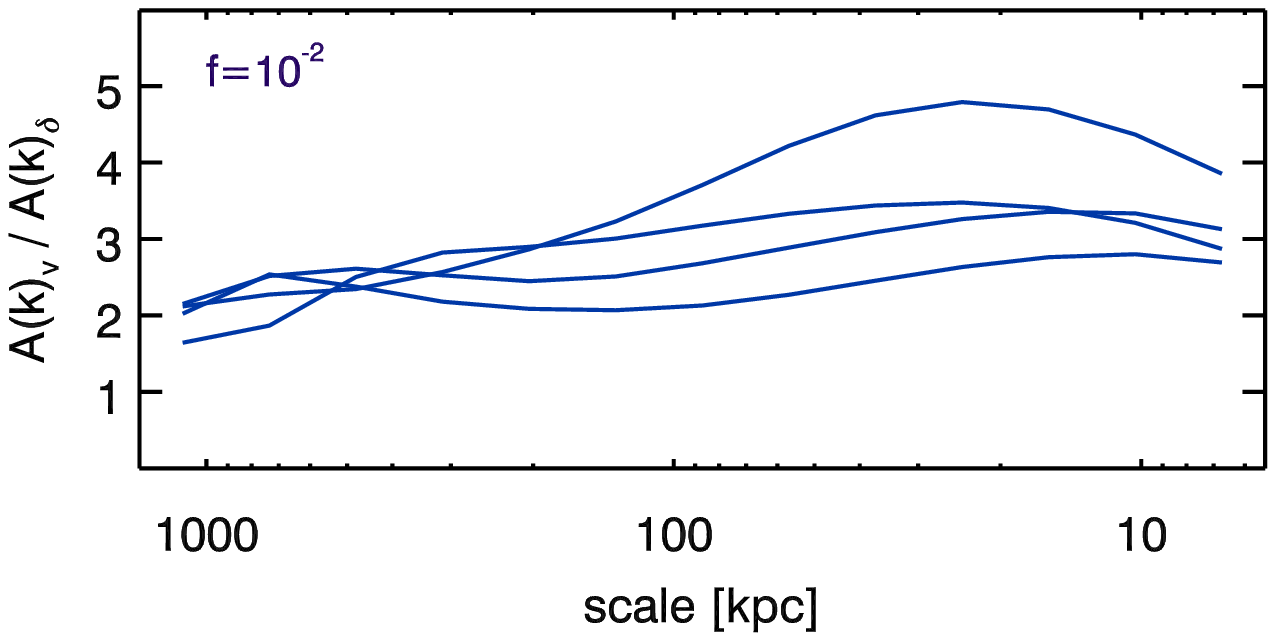}}
       \subfigure{\includegraphics*[scale=0.54]{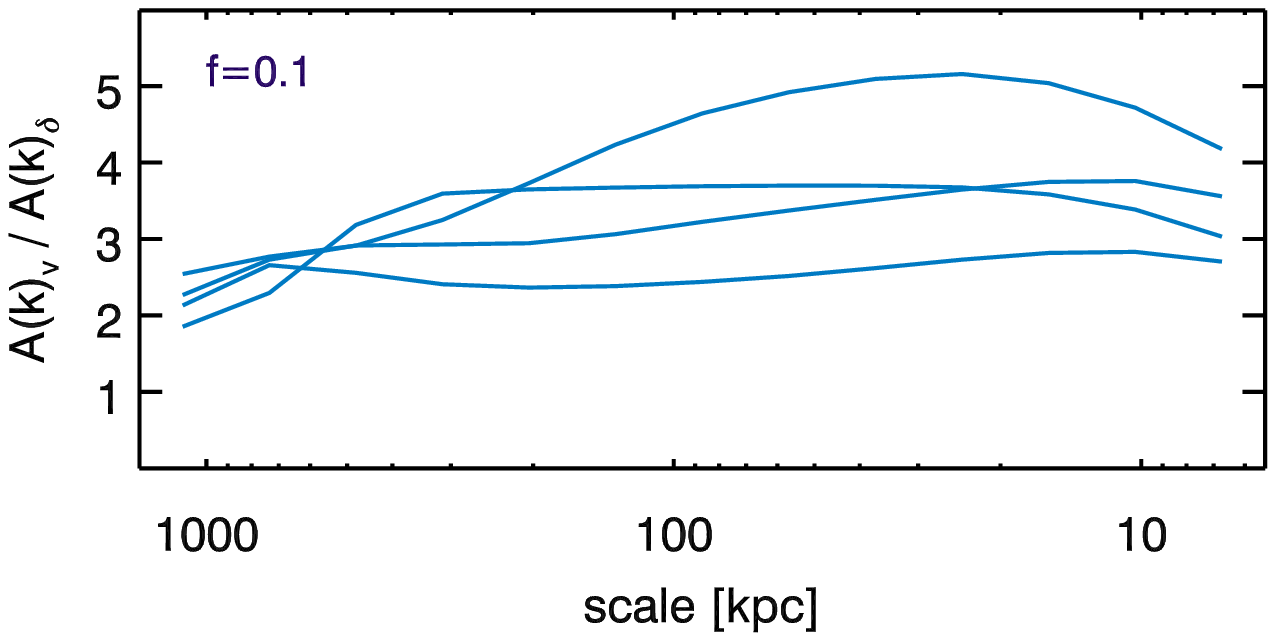}}
       \subfigure{\includegraphics*[scale=0.54]{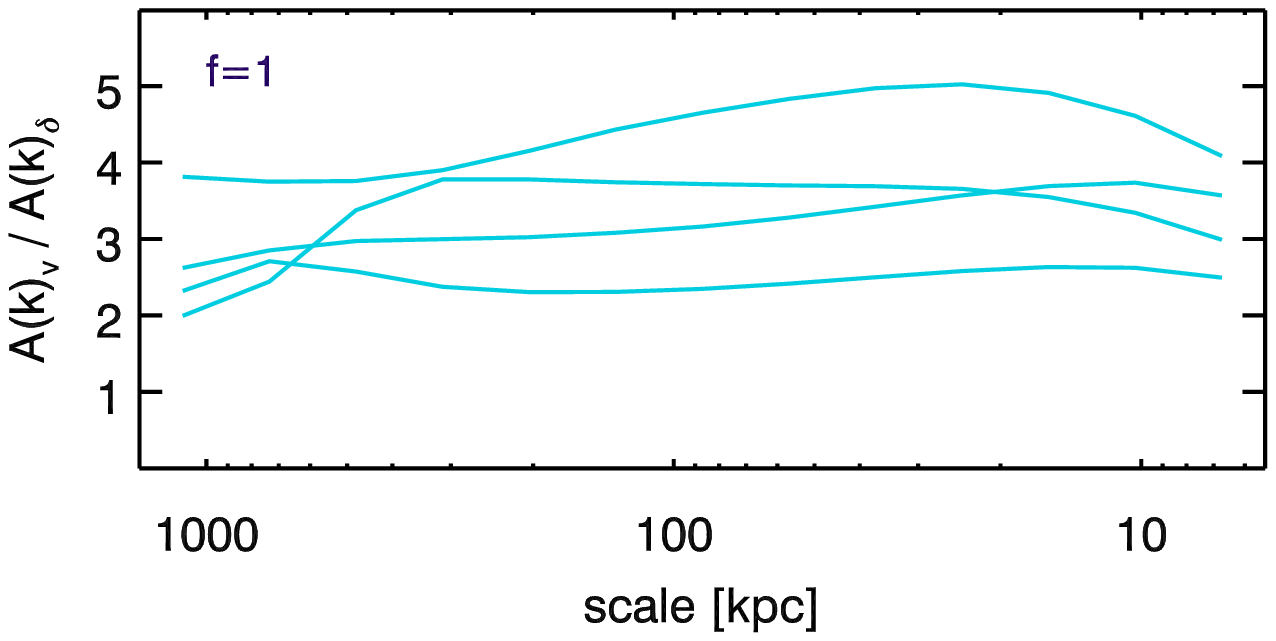}}                                     
       \caption{Ratio of the power spectrum related to total velocity and density perturbations, for all the computed models. Each panel groups the models with identical conductivity, but different $M$ (same colors as in Fig.~\ref{fig:Ak}). $A_\rho$ for $L/2$ runs is rescaled by a factor $2^{1/3}$, to emphasize slope similarities. In the hydro runs, the ratio is very tight, decreasing from $\sim\,$2.5 to roughly unity, near the dissipation scale. An increasing ratio instead marks the presence of significant conduction. Lower Prandtl numbers ($P_{\rm t}\propto M/f$) lead to wider scatter, with ratios up to $\sim\,$5. 
       The $A_v/A_\rho$ ratio is a new key diagnostics able to unveil the presence of substantial conductivity in the ICM.}        
     \label{fig:ratio}
     \end{center}
\end{figure}

In Figure \ref{fig:Ak}, we show the retrieved characteristic amplitude of $v/c_{\rm s}$ (blue; notice that $A_v=\sqrt{A^2_{v_{\rm x}}+A^2_{v_{\rm y}}+A^2_{v_{\rm z}}}$),
superposed to the amplitude of density perturbations (red), after reaching statistical steady state ($\gta2\,t_{\rm turb}$). 
From the top left panel, the models have increasing turbulence: weak ($M\sim0.25$), mild ($M\sim0.5$), and strong ($M\sim0.75$). The ratio of turbulent to thermal energy is 3.5, 14 and 31 percent ($E_{\rm turb} \simeq 0.56\, M^2 E_{\rm th}$), respectively.
The last model (bottom right panel) tests weak turbulence with half the reference injection scale ($\sim\,$300 kpc). 
The global behavior of the spectra related to density, velocities, and their ratio (Fig.~\ref{fig:ratio}) is fairly self-similar,
over different Mach numbers and injection scales.
We covered in GC13 the details of the $\delta\rho/\rho$ spectrum,
we focus here on the turbulent velocity and their relationship.

The normalization of the velocity spectrum sets the level of perturbations, tied to the $A(k)$ peak. 
In Figure \ref{fig:ratio}, we better illustrate the $A_v(k)/A_\rho(k)$ ratios. The hydro $f=0$ runs (top)
show a converging maximum value $A_v/A_\rho\approx2.3$ (using the 1D velocity $A_{v_{\rm 1D}}/A_\rho\approx 1.3$), implying that stronger turbulence linearly induces larger density fluctuations. This is a key result that allows to quickly estimate ICM perturbations via the leading turbulent motions, and vice versa. For low $M$, gravity waves mainly produce entropy perturbations as $\delta K/K \propto v_{\rm 1D}$ (see \S\ref{s:gw}). For slow motions, the isobaric mode is respected, hence $v_{\rm 1D} \propto \gamma\, \delta \rho/\rho$, as also simulated. For $M\gta0.5$ (see \S\ref{s:pw}), compressive sound waves start to significantly contribute: entropy perturbations remain constant, while $\delta P/P$ increases, sustaining the same linear relation, but smoothly shifting towards the adiabatic mode. 
In section \ref{s:disc}, we thoroughly discuss all the thermodynamic perturbations (Fig.~\ref{fig:M025_all}$\,$-$\,$\ref{fig:Mhigh_all}) and the underlying physical interpretation.

It is important to note that, for applications in other studies, proper attention should be paid to the conversion to adopt, given the initial quantity or observable. Shifting from the spectral to physical integrated\footnote{The total variance can be computed integrating $P(k)\,4\pi k^2 dk$ over the whole range of scales. The total variance (as the $M$ value) is here typically $\sim1.7$ times higher than the $A(k)$ peak.} quantities, the velocity/density ratio  remains the same. 
However, if we relate the real-space Mach number (i.e.~the total variance) to the spectral peak, 
as in GC13, the conversion to use is $M \approx 4\, A(k)_{\rho,{\rm max}}$ (assuming $L\sim600$ kpc). 
Furthermore, turbulence can not create relative density perturbations with amplitude higher than roughly the 1D Mach number (\S\ref{s:disc}).
If significantly violated, this would indicate that the perturbation or velocity field has been contaminated by 
the unfiltered background profile or laminar flows (which are particularly complex in unrelaxed systems). 
Similarly, strong inhomogeneities must be properly removed: the linear relation applies to relatively small perturbations $\delta \rho/\rho < 1$, not to features as cold fronts or buoyant bubbles.

\begin{figure} 
    \begin{center}
       \subfigure{\includegraphics[scale=0.55]{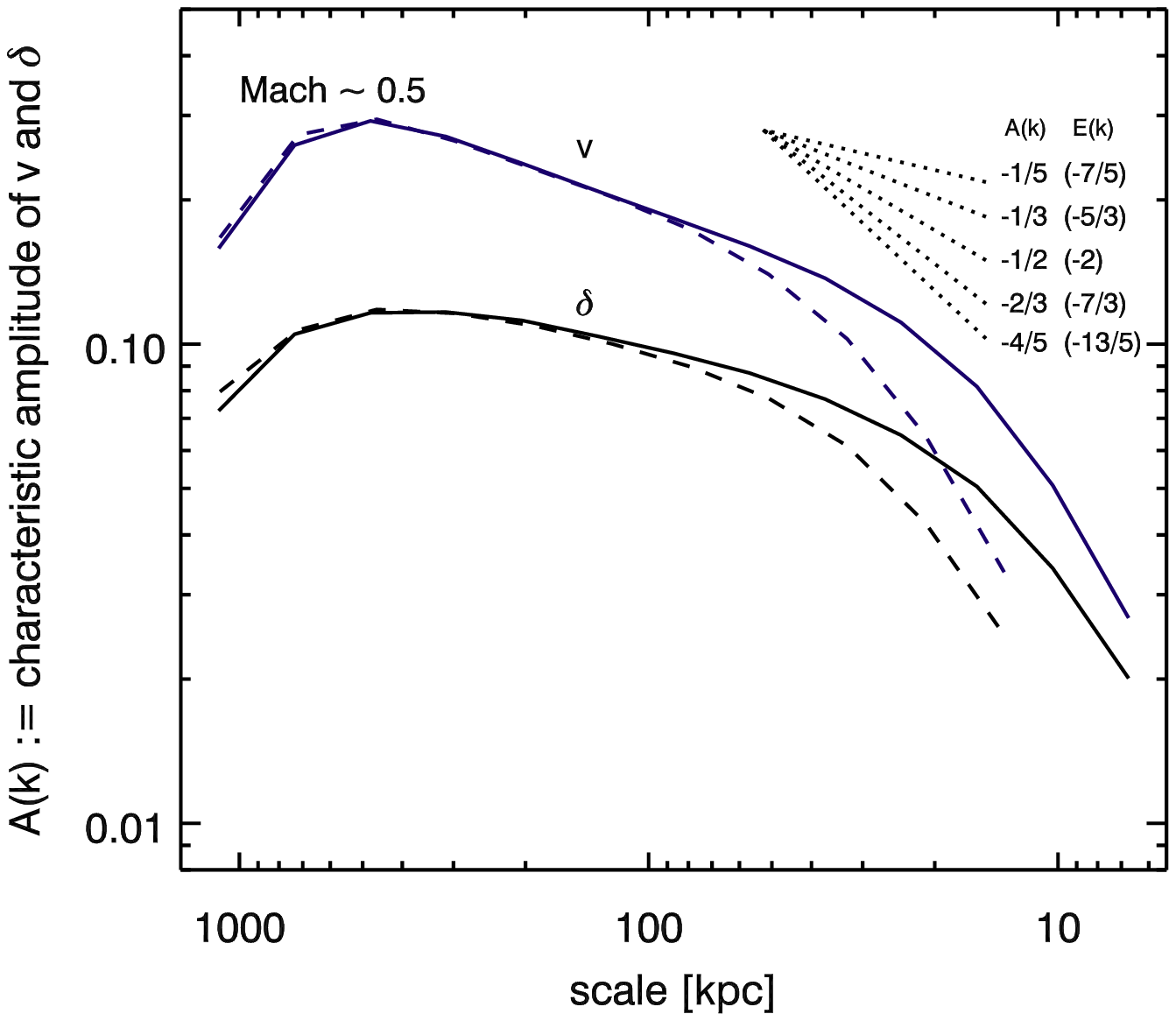}}
       \subfigure{\includegraphics[scale=0.55]{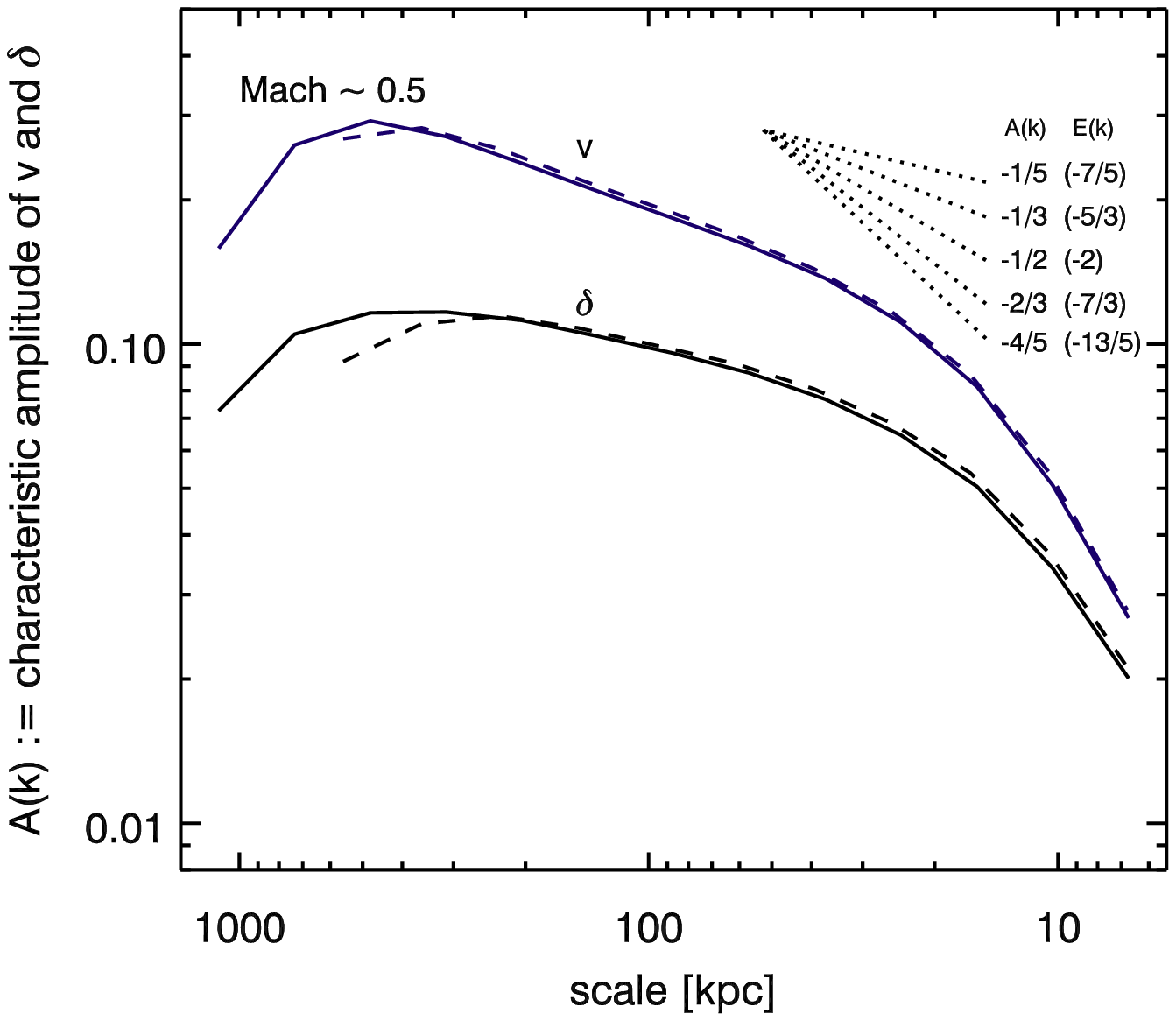}}
\caption{Top: Characteristic amplitude of $v/c_{\rm s}$ and $\delta \rho/\rho$, for the hydro model with $M\sim0.5$, doubling the numerical viscosity, i.e.~using $2\times$ lower resolution (similar to the Spitzer value; dashed lines). Spectra are convergent, except at small scales where the increased numerical diffusivity damps both $v$ and $\delta\rho/\rho$ at $\sim$2 times the original dissipation scale. The density cascade is affected in a slightly different way by the larger diffusivity; some deviations from the classic advection theory of tracers are thus expected (\S\ref{s:dep}).
       Bottom: Spectrum of the above model extracted in the full box (solid) and in the center (dashed; $< r_{500}/4$), where stratification is less prominent.
       Turbulence and density perturbations are overall homogeneous, despite the cluster stratification (cf.~Fig.~\ref{fig:maps}). 
      }       
        \label{fig:core}
     \end{center}
\end{figure}

Below the injection scale, $A_v$ and $A_\rho$ continue to be tightly related in the non-diffusive models (Fig.~\ref{fig:ratio}).
The ratio is independent of $M$, as indicated by the tight scatter.
It is remarkable that the density acts as effective `tracer' of the velocity field, developing a similar inertial cascade. This is also true for the leading entropy/pressure perturbations. The phenomenon can be explained via the classic theory of advection of passive tracers in turbulent media, $A(k)_\rho \propto A(k)_v$ (\citealt{Obukhov:1949,Corrsin:1951}; see \S\ref{s:cascade}).
On the other hand, we observe that, in the hydro runs, the ratio steadily declines as $l^{0.13}$, reaching about unity near the dissipation scale. The decrease is associated with a shallower cascade of perturbations, due to the initial radial gradients, compressive features, and differences in the diffusivity of the `tracer' (\S\ref{s:dep}). In Figure \ref{fig:core} (top), we show a test with $2\times$ lower resolution (i.e.~$2\times$ higher effective viscosity). Aside the good large-scale convergence, the run clarifies that the density/tracer is susceptible to diffusivity in a slightly different way compared with $v$,
hence we expect departures from the classic tracers theory. 
Notice also how the cascade of density perturbations is not a perfect power law, but tends to exponentially decline, even in the hydro run.

Using half the reference injection scale, the $A_v$ peak is analogous to the that of the reference run (conserving $M\sim0.25$),
while the density perturbations slightly decrease by $\sim\,$$2^{1/3}$, i.e.~the previous cascade truncated at $L/2$.
Stirring smaller scales reduces 
the influence of gravity waves, since the zone where the turbulence frequency ($\propto L^{-1/3}$) is shorter than the 
buoyancy frequency shrinks (Fig.~\ref{fig:freq}). 
By varying the injection scale between $L'=L\equiv600$ kpc, $L'=L/2$, and $L'=L/3$ we find that correcting the ratio by a factor $\sim (L'/L)^{1/3}$ restores the normalization to a universal value (Fig.~\ref{fig:ratio}, top).
Also, the development of the shorter cascade is hindered by the progressive proximity to the dissipation scale.

A key result is the substantial decoupling of velocities and density perturbations, as we increase the level of conduction (dark to bright line color: $f=0\,, 10^{-3},\, 10^{-2},\, 10^{-1},\, 1$; Fig.~\ref{fig:Ak}$\,$-$\,$\ref{fig:ratio}). 
In other words, the quick transfer of heat damps density fluctuations (the forming overdensities quickly re-expand due to the temperature increase), while it leaves unaltered the turbulent velocity cascade, or momentum transfer. The rising $A_v/A_\rho$ ratio (Fig.~\ref{fig:ratio}, top to bottom) is a crucial result that provides a new constraint on the conductive state of the ICM, in addition to the slope of the spectrum. It also breaks any minor degeneracy that strong conduction ($f\gta0.1$) may induce in the spectrum, due to the global damping of power (slightly flattening the cascade). The upcoming {\it Astro-H} mission will provide important constraints on the ICM turbulent velocities; combined with high-quality determinations of density perturbations via {\it Chandra/XMM} (and {\it Athena}), our knowledge of the ICM physics could significantly improve, exploiting the $A_v/A_\rho$ diagnostic.

The velocity cascade follows the Kolmogorov index ($A(k)\propto k^{-1/3}$ or $E(k)\propto k^{-5/3}$) in all runs, except with weak turbulence, where it becomes slightly steeper, though with increased scatter. 
Considering the substantial stratification, it is remarkable that classic Kolmogorov theory consistently applies to a cluster atmosphere (see also \citealt{Vazza:2011,Valdarnini:2011}).
The density spectrum instead displays
a steep decay towards $A_\rho\propto k^{-1/2}$ as the turbulent Prandtl number $P_{\rm t}\equiv t_{\rm cond}/t_{\rm turb}\lta100$ (see GC13\footnote{For the weak turbulence run with $f=1$, $P_{\rm t}\sim1$ at $L=600$ kpc. We note $P_{\rm t}$ can be also seen as a turbulent Peclet number, if turbulence is interpreted as an advection -- rather than diffusion -- process.}).
Such a steepening induces $A_v/A_\rho$ to become gradually shallower\footnote{In general, significant diffusivity (especially numerical) acting on both $\rho$ and $v$ tends to align the two spectra, even on small scales, a common feature we found in cosmological simulations (Z14 and Fig.~\ref{fig:core}).},
inverting the trend for $f\gta10^{-2}$ (Fig.~\ref{fig:ratio}, third panel). This can be explained in terms of the advection theory of tracers (\S\ref{s:cascade}), where the diffusivity of the scalar has no effect on the velocity cascade.
The scatter of $A_v/A_\rho$ rises with decreasing Prandtl number, i.e.~with stronger 
conduction and weaker turbulence ($P_{\rm t}\propto M/f$).
Conductivity with $f\gta0.1$ can globally stifle the regeneration of perturbations by a factor of 2$\,$-$\,$4. 
Fig.~\ref{fig:ratio} confirms that any substantial conductivity in the ICM will clearly emerge in the $A_v/A_\rho$ diagnostic, showing values up to $\approx\,$5 and 3 for weak and strong turbulence, respectively, even at scales of 100s kpc.

\section[]{Real-space properties and X-ray constraints} \label{s:maps}  

\begin{figure*}%[htb]
    \centering
       \subfigure{\includegraphics[scale=0.4786]{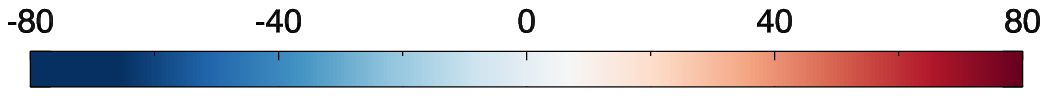}}
       \subfigure{\includegraphics[scale=0.4786]{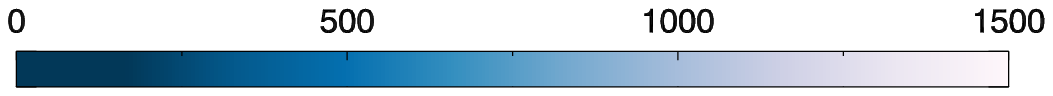}}
       \subfigure{\includegraphics[scale=0.4786]{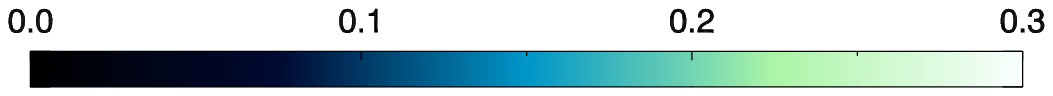}}
       \subfigure{\includegraphics[scale=0.4786]{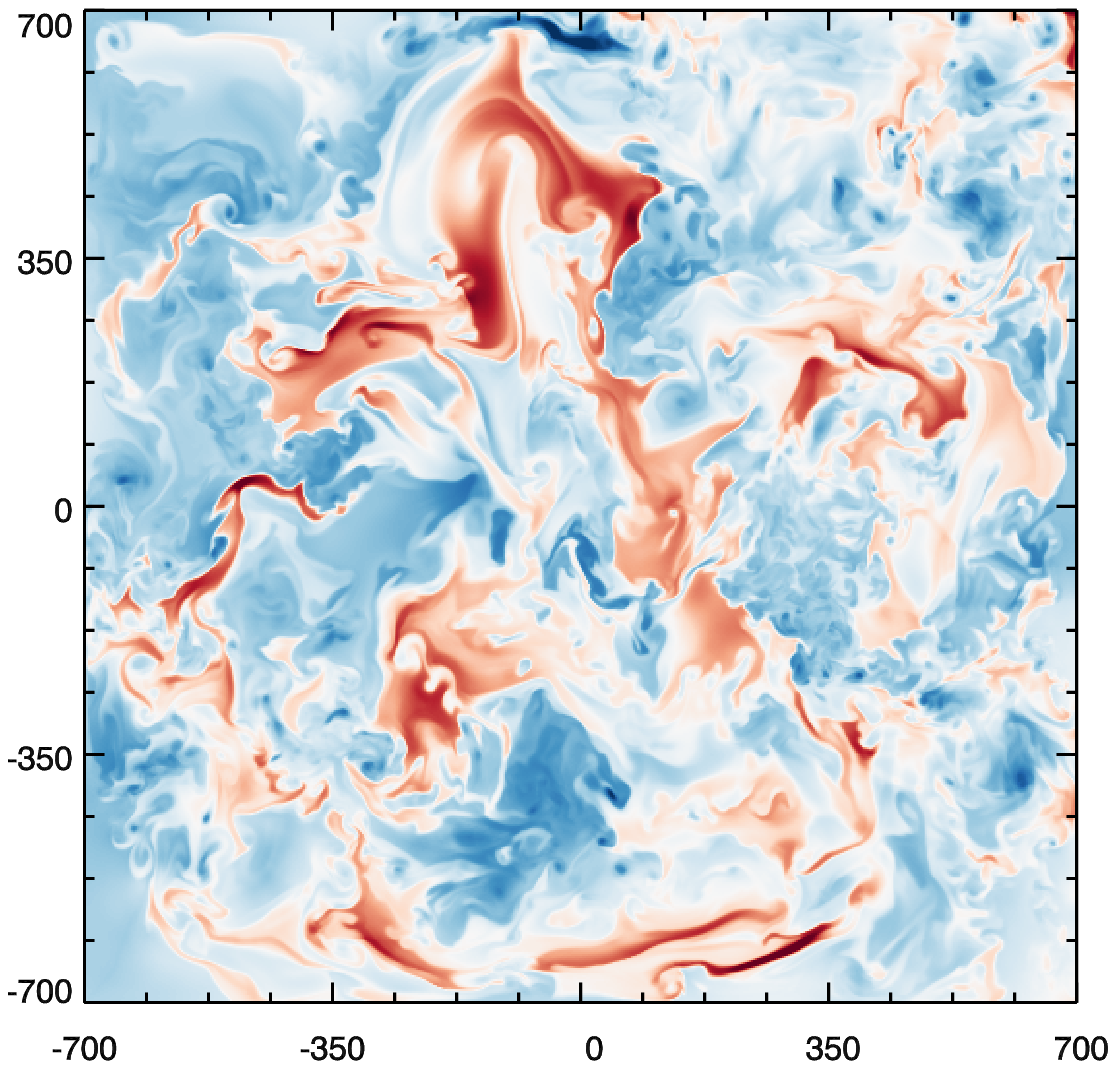}}                     
       \subfigure{\includegraphics[scale=0.4786]{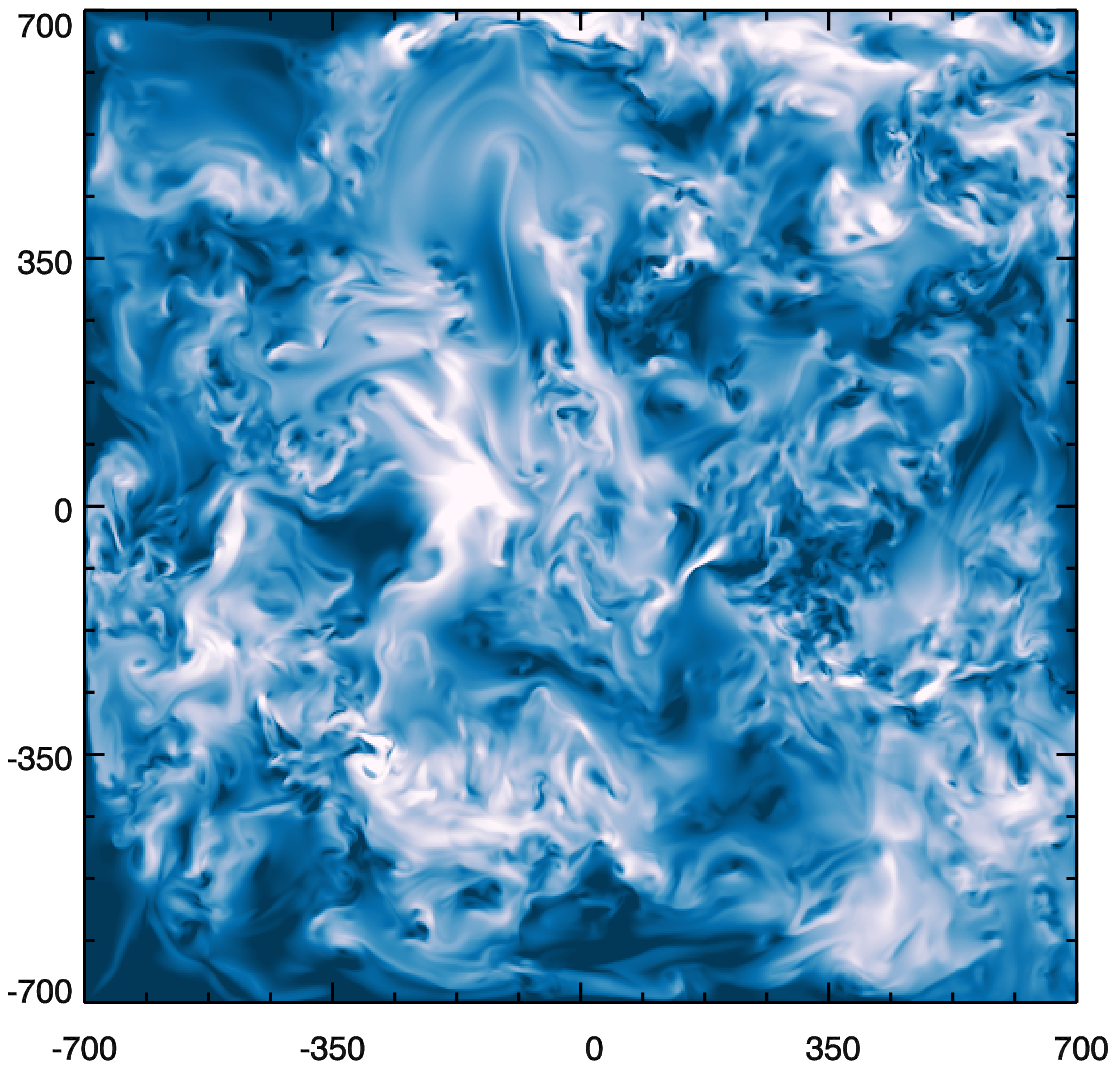}}
       \subfigure{\includegraphics[scale=0.4786]{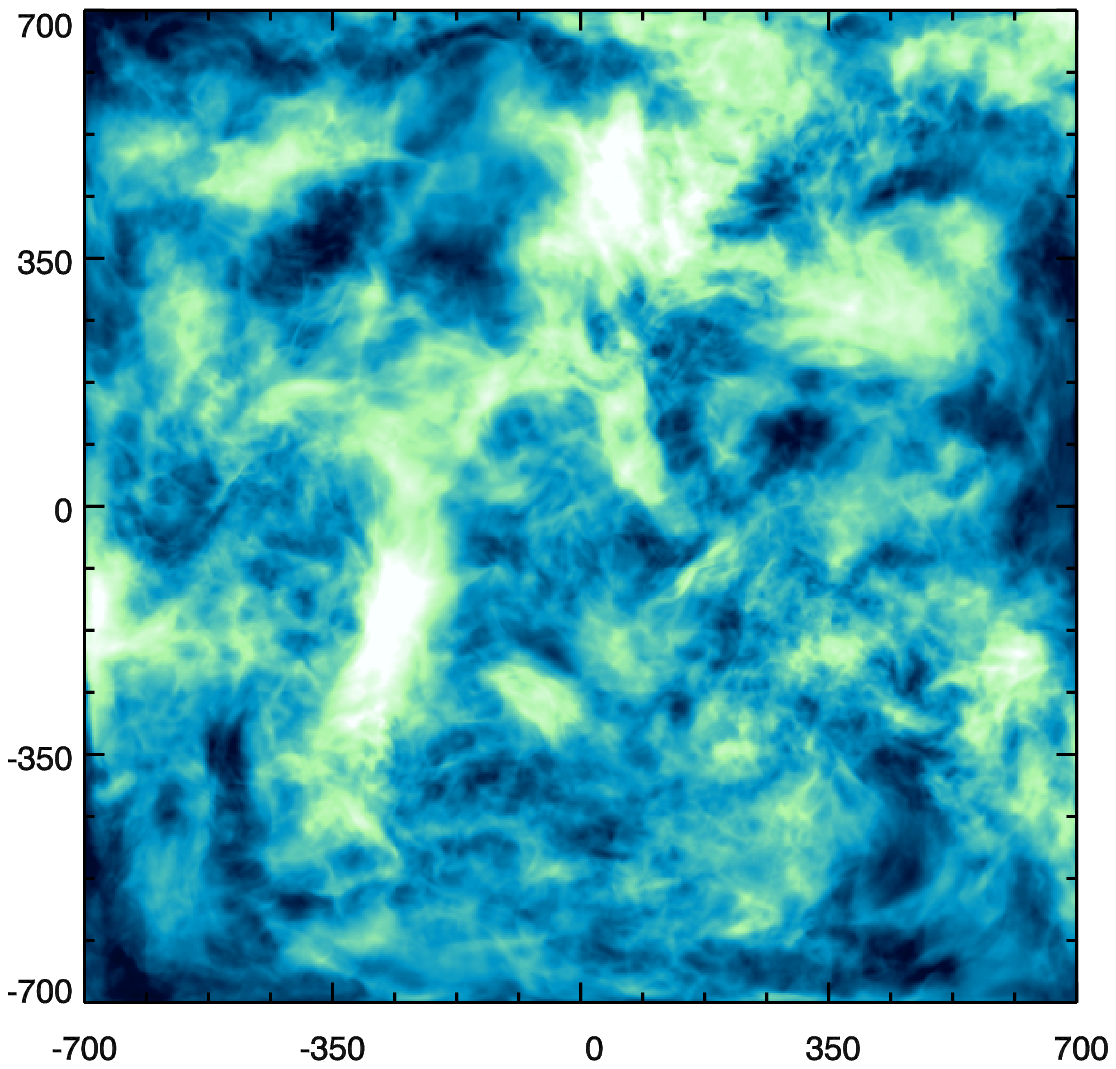}}                           
       \subfigure{\includegraphics[scale=0.4786]{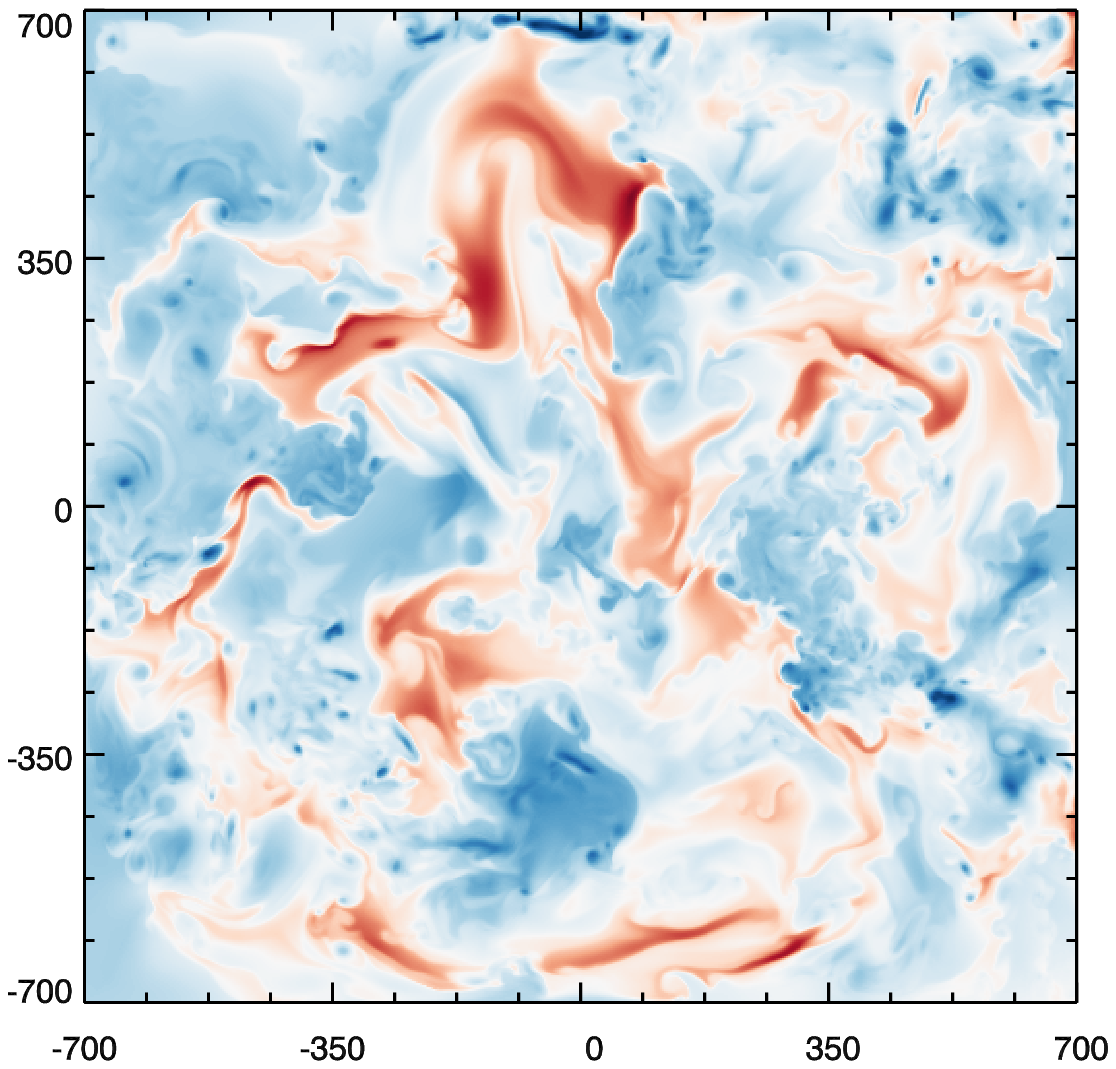}}  
       \subfigure{\includegraphics[scale=0.4786]{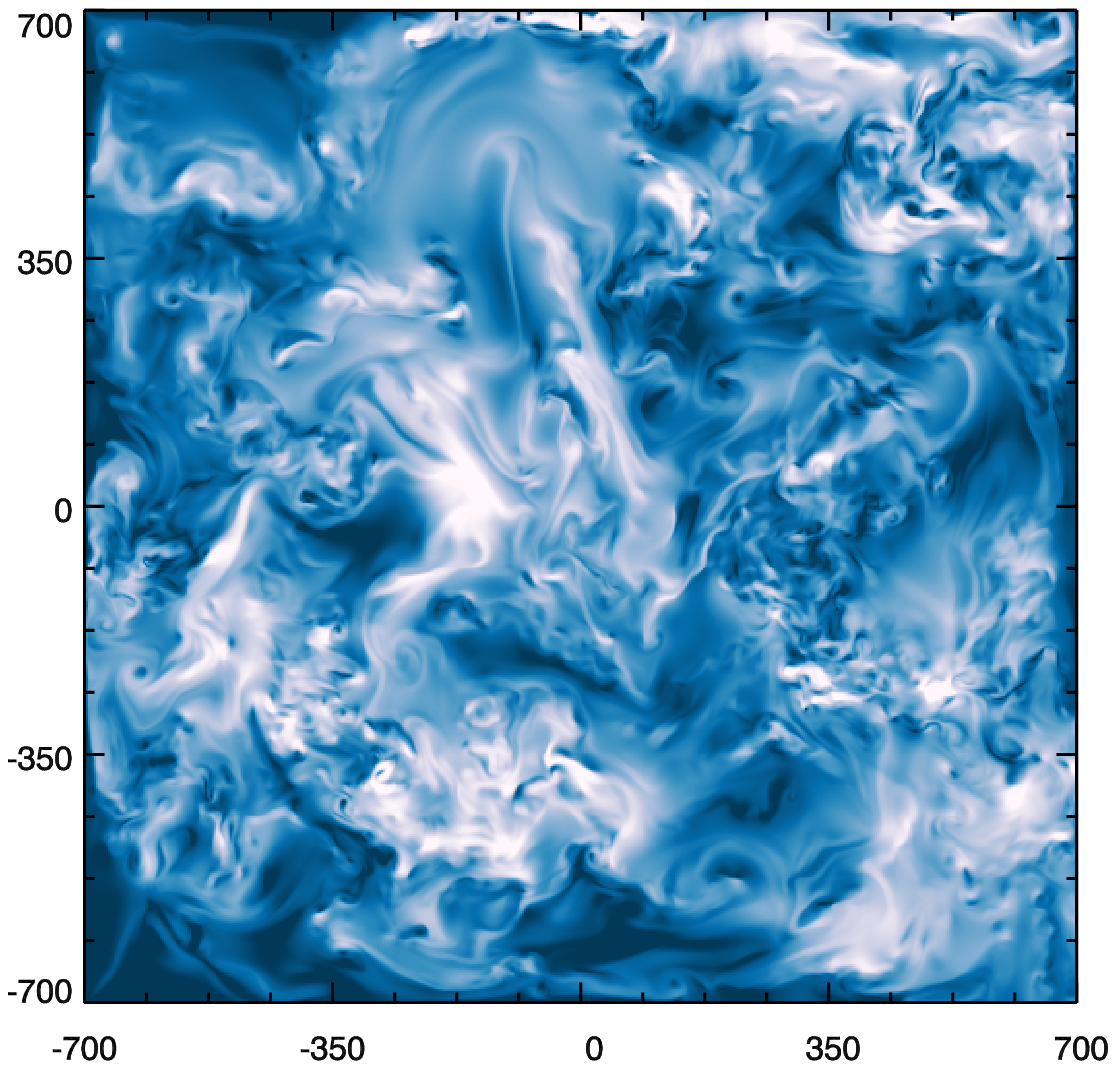}}
       \subfigure{\includegraphics[scale=0.4786]{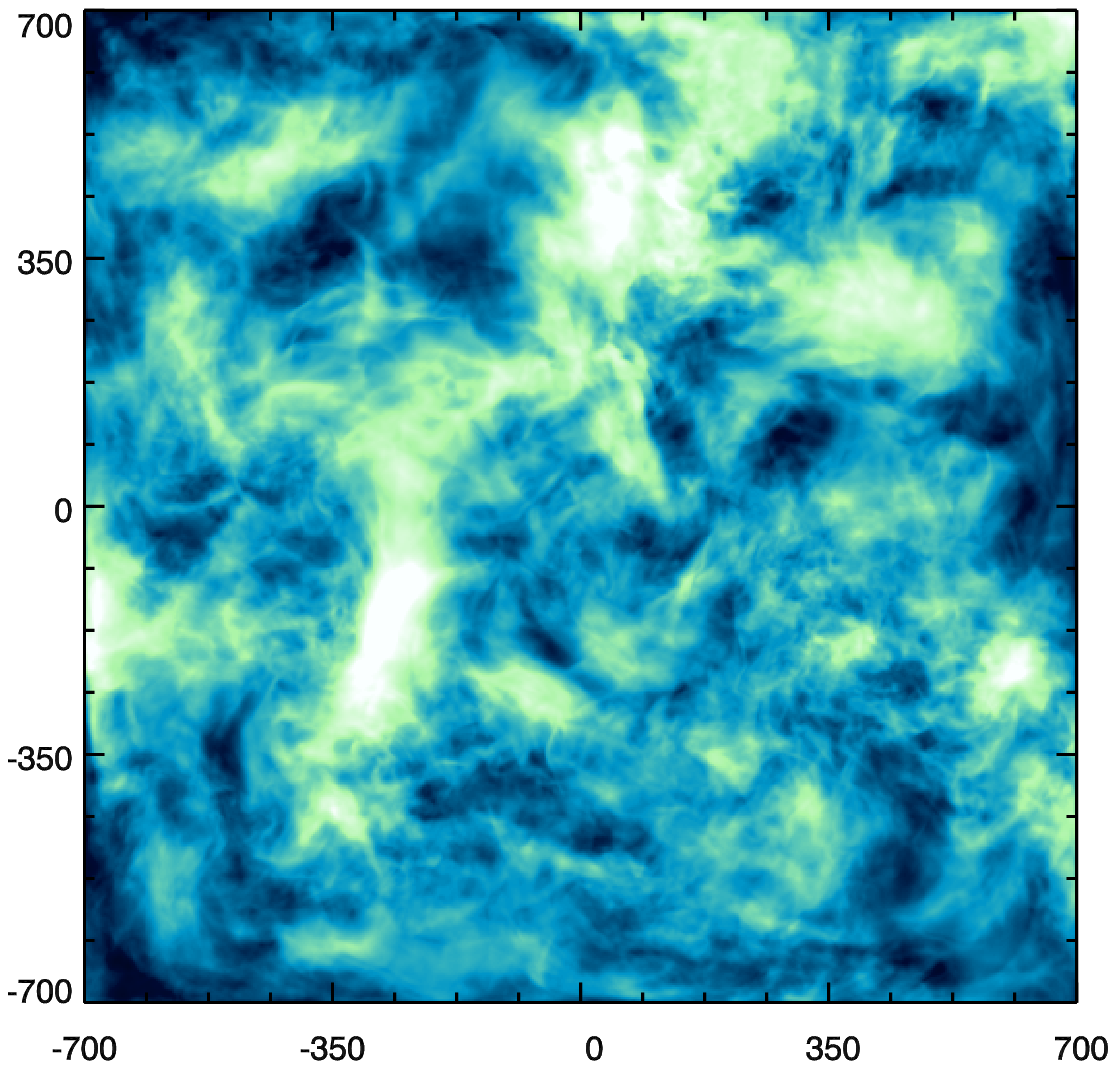}}                          
       \subfigure{\includegraphics[scale=0.4786]{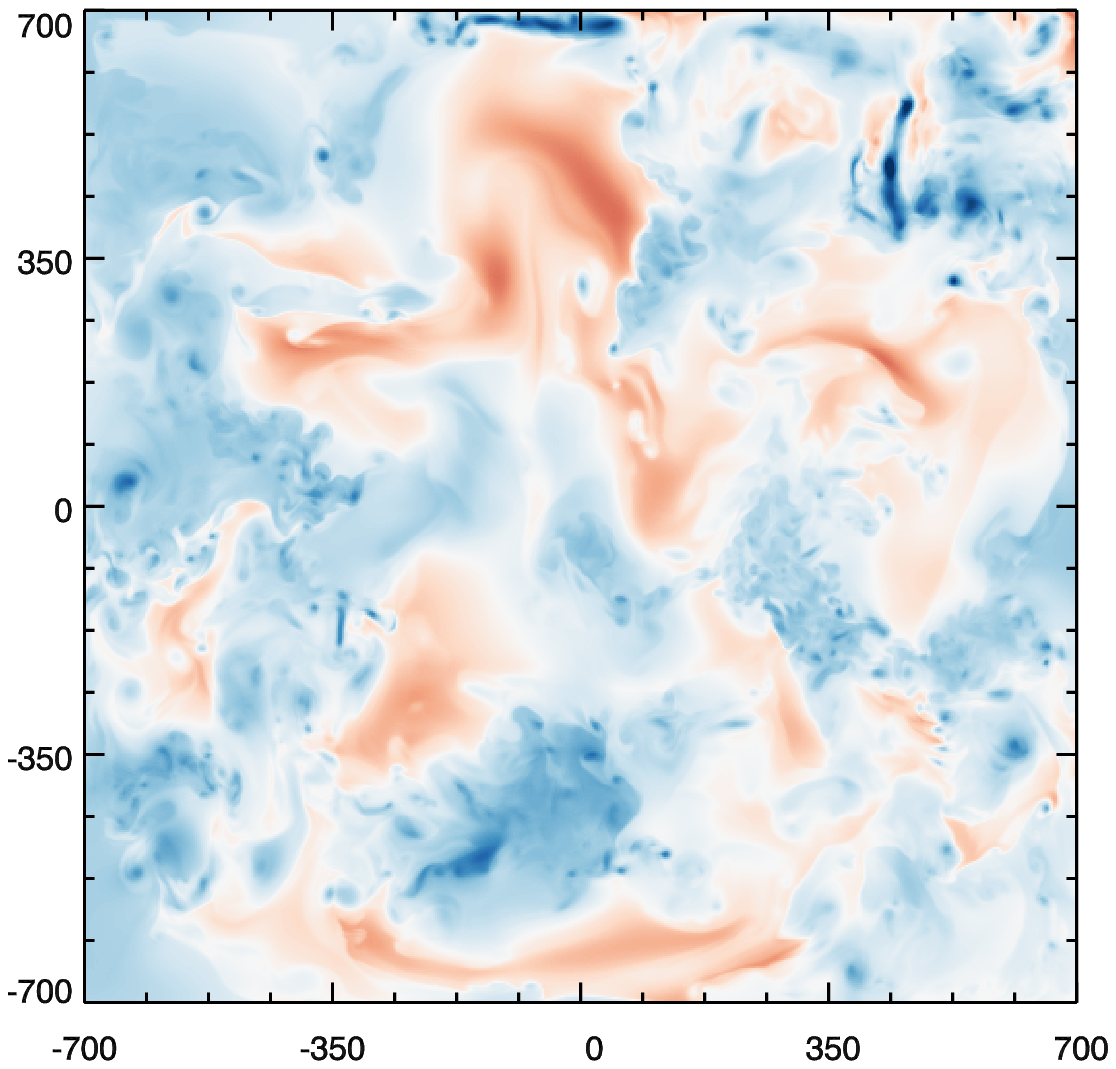}}
       \subfigure{\includegraphics[scale=0.4786]{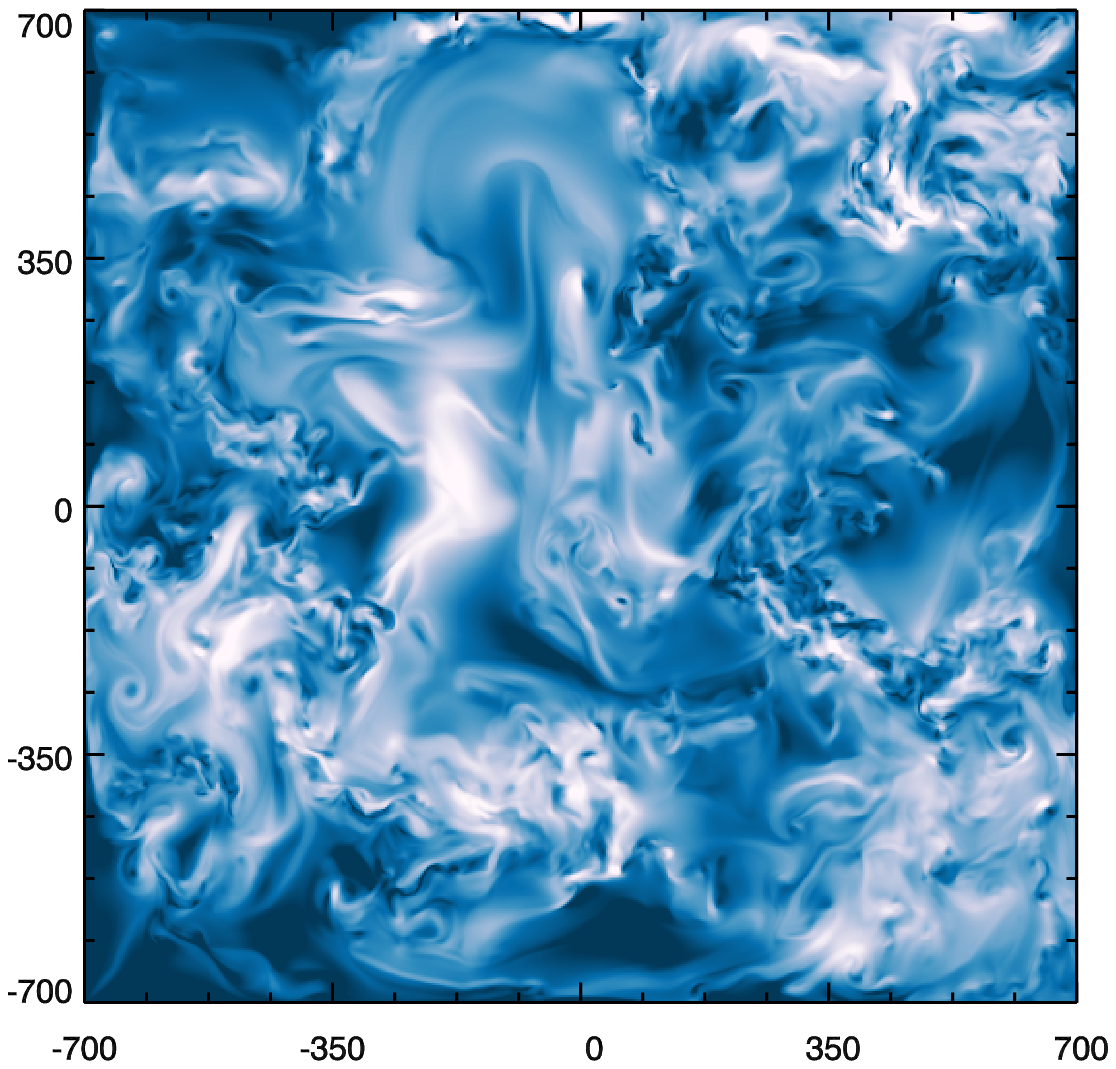}}
       \subfigure{\includegraphics[scale=0.4786]{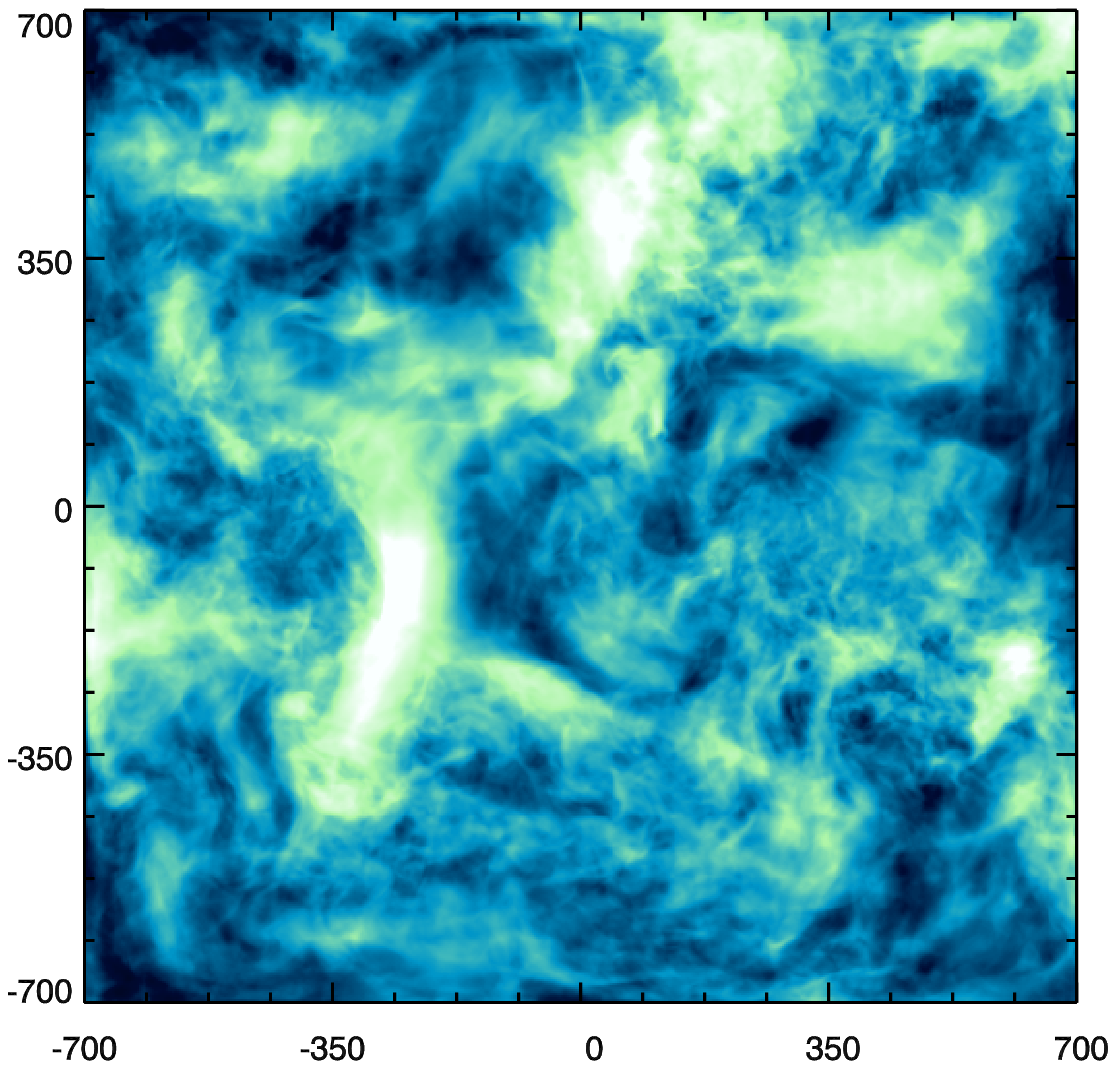}}                   
       \subfigure{\includegraphics[scale=0.4786]{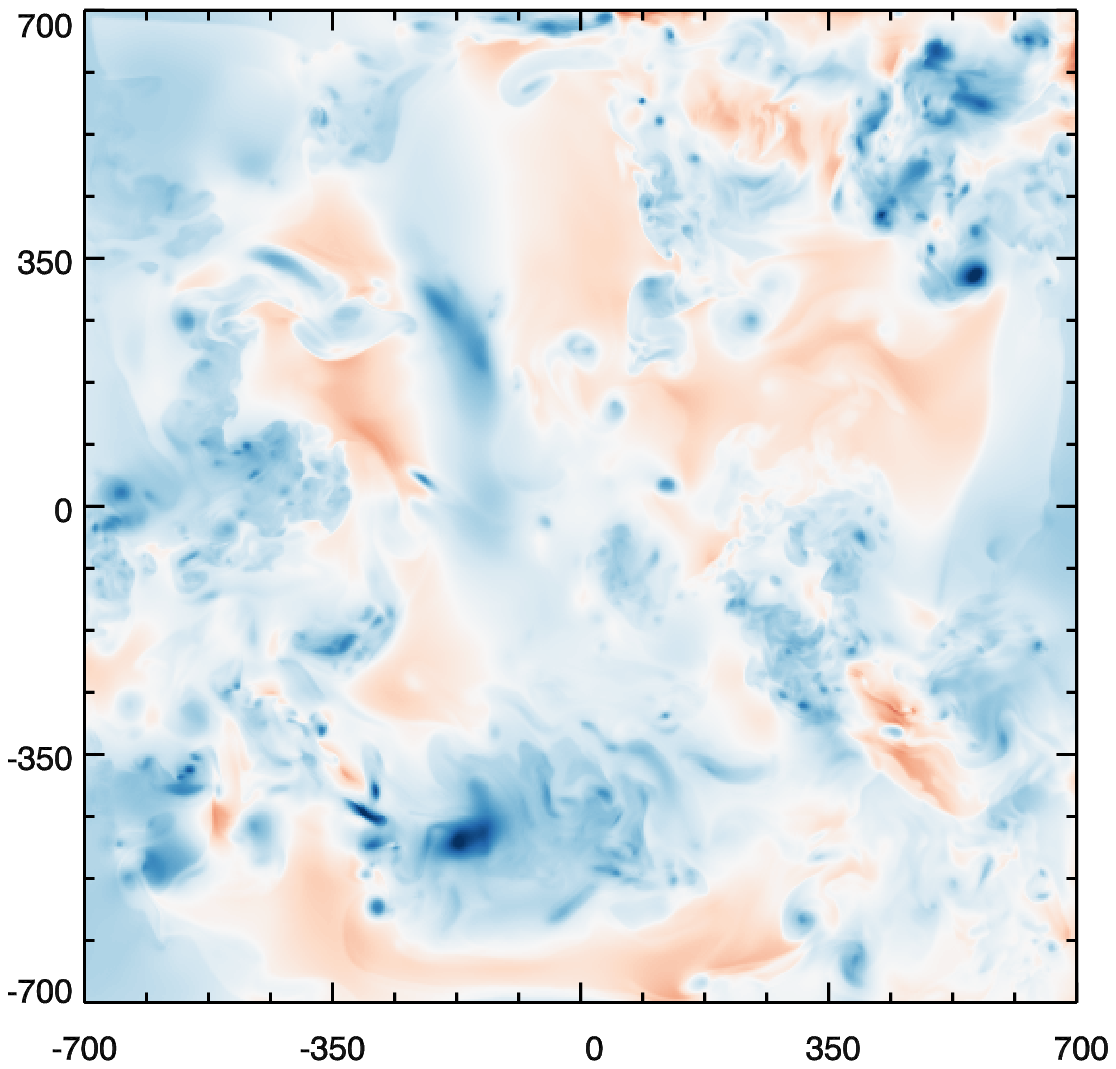}}
       \subfigure{\includegraphics[scale=0.4786]{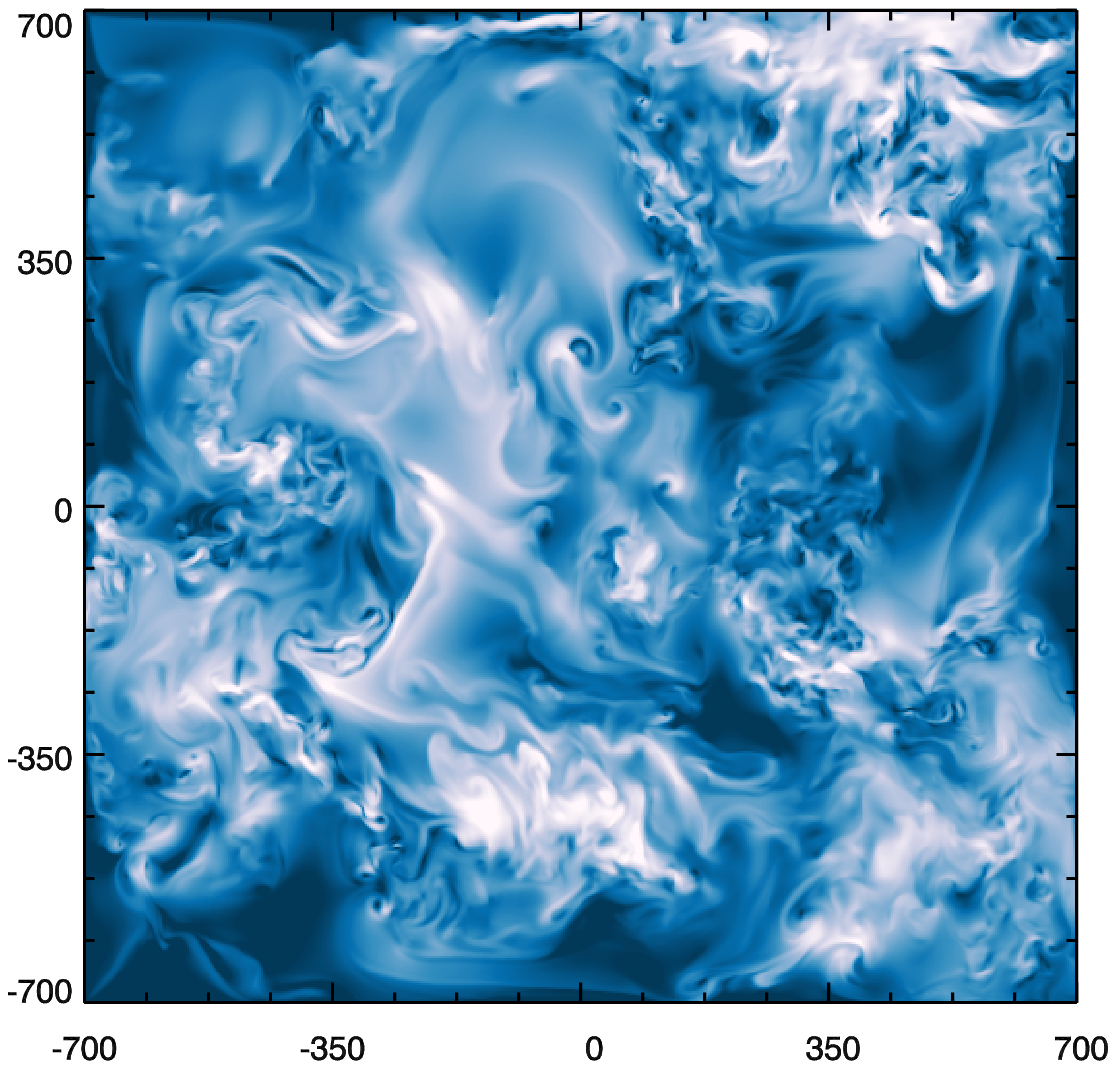}}
       \subfigure{\includegraphics[scale=0.4786]{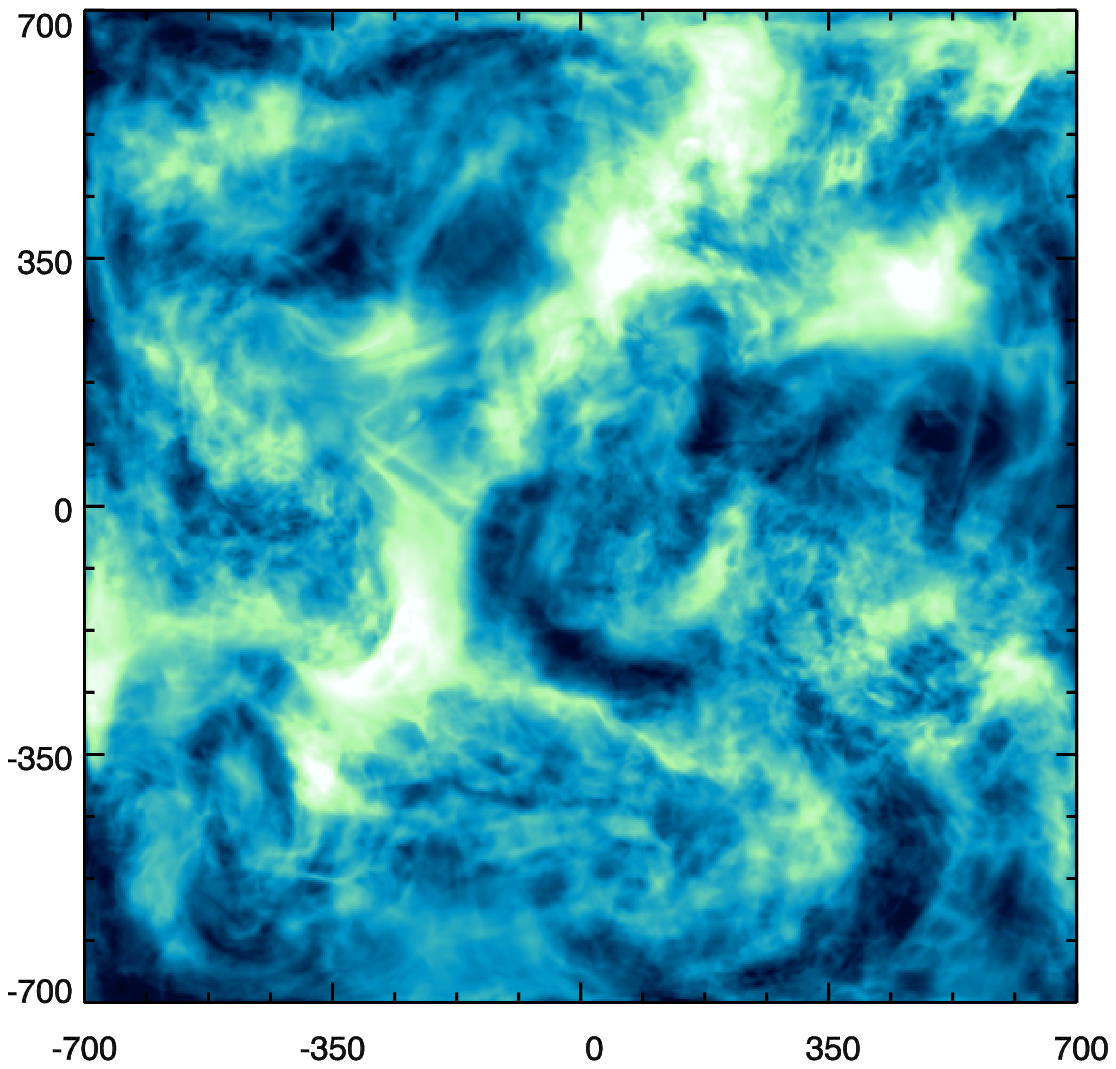}}                   
       \caption{Left: Mid-plane cross-sections of $\delta \rho/\rho$ (per cent) for the models with $M\sim0.5$. From top to bottom: increasing conduction with $f=0\ {\rm (hydro)},\, 10^{-3},\, 10^{-2},\, 10^{-1}$ (the latter very similar to the $f=1$ run). Middle: Same cross-sections but for the module of total velocity (km s$^{-1}$).           
         The hydro runs show sharp filamentary density structures produced by the turbulent velocity field and later deformed by Kelvin-Helmholtz and Rayleigh-Taylor instabilities. 
       Strong conduction damps instead these perturbations, while leaving
       unaltered the Kolmogorov cascade of turbulent eddies. The $v$ and $\delta\rho/\rho$ fields have correlated amplitude, but different phases (the density field is the tracer).
         Right: Observed relative line broadening (per cent) due to turbulent motions along the $y$-axis view, $\Delta E/E_0 \equiv \sqrt{2}\, \sigma_{\rm 1D, ew}/c$,
         where $\sigma_{\rm 1D,ew}$ is the projected X-ray emission-weighted velocity dispersion. 
       The forthcoming {\it Astro-H} telescope will be able to detect projected turbulent velocities above $\sim200$ km s$^{-1}$, i.e.~$\Delta E/E_0>0.1$ per cent (FWHM$\,=1.66\,\Delta E$), using the Fe XXV line. 
     }
     \label{fig:maps}
\end{figure*} 

Before delving into the theoretical interpretation (\S\ref{s:disc}), it is worth to understand the real-space properties related to turbulence and perturbations of density (or the `tracer'), and what X-ray observations can detect through
the spectral line broadening and the projected images.
As reference, we consider the models with $M\sim0.5$ ($E_{\rm turb}/E_{\rm th}\sim0.14$, a common cluster regime; e.g.~\citealt{Schuecker:2004, Lau:2009}; GC13 -- sec.~4.3). 

In Figure \ref{fig:maps}, we compare the mid-plane cross-sections of $\delta\rho/\rho$ (left) and 
magnitude of total velocity (middle).
The key result is the progressive smoothing of density fluctuations raising the level of conduction, while the turbulent velocity field remains unaltered  (top to bottom panels: $f=0,\, 10^{-3},\, 10^{-2},\, 10^{-1}$). 
In the hydro run, the perturbation field shows a complex morphology of filamentary structures, 
produced by the turbulent velocity field
and later deformed by Kelvin-Helmholtz and Rayleigh-Taylor instabilities.
The rolls and filaments are almost washed out in the presence of strong conduction ($f\gta0.1$),
transforming the perturbations from isobaric to isothermal (\S\ref{s:gw}).
On the other hand, strong conductivity can not completely wipe out fluctuations (bottom maps). While
entropy fluctuations decrease, compressive pressure perturbations still maintain the same level (Fig.~\ref{fig:Mhigh_all}, bottom). In other words, strong conduction can also promote minor fluctuations,
due to the fast transfer of heat and change of compressibility in the medium.

All the velocity maps (middle) are remarkably similar, both statistically and locally, with minute differences only if $f\gta0.1$.
As density fluctuations, the turbulent velocities do not show any major difference within or outside the cluster core, signaling a significant level of homogeneity. 
To be more quantitative, we extracted the spectra only from the cluster center ($r < r_{500}/4$),
where stratification is less prominent.
As shown in Fig.~\ref{fig:core} (bottom), both the $\delta\rho/\rho$ and velocity spectra are similar to those computed in the full box.
At the largest scale, $\delta\rho/\rho$ experiences a minor decline ($\sim$10\%), in part because the entropy/pressure profile is shallower in the core, in part due to the limited statistics related to the smallest modes.
Concerning isotropy, the real-space maps also do not highlight major deviations. However, transforming the velocity field in spherical coordinates, we retrieve mild anisotropies (Fig.~\ref{fig:beta}). As $M<0.5$, the large-scale velocities become slightly more tangential, due to the stronger influence of stratification (\S\ref{s:interp}). Instead, stronger turbulence (and conduction) increases the relevance of sound waves, restoring isotropy.

Fig.~\ref{fig:maps} points out that
the phases of the perturbation field are not coincident with that of the velocity field, although strongly correlated in amplitude. 
Pearson coefficient indicates a negligible anticorrelation between $\delta\rho/\rho$ and total velocity ($R < -0.2$), in all models. In other words, the density filaments are not strictly tied to a local high velocity.
This is expected since 
the cause of fluctuations is the turbulence driving, while $\delta\rho/\rho$ plays the
role of the tracer in a continuously chaotic environment.

\begin{figure} 
       \centering
       \subfigure{\includegraphics[scale=0.52]{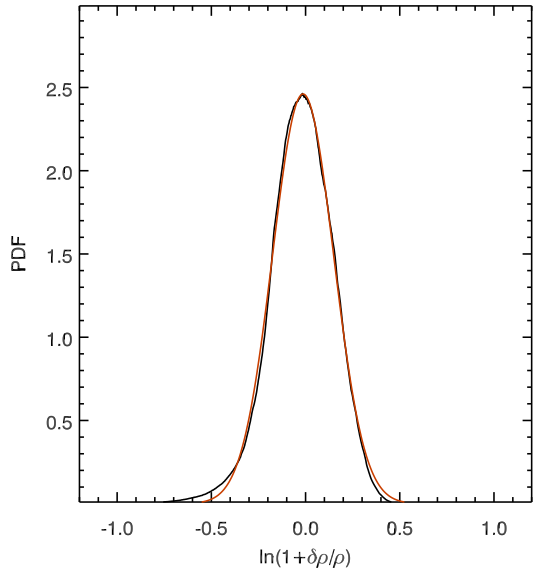}}
       \caption{Volumetric PDF of the logarithmic density fluctuations, $\ln(1+\delta\rho/\rho)$, for the run with $M\sim0.5$ and $f=0.1$ (black). The thermodynamic fluctuations can be in general described by a log-normal distribution (red line), with small corrections due to high-order moments (skewness and kurtosis).
       }
     \label{fig:PDF}
\end{figure}

We analyzed the volumetric PDF of the logarithmic density fluctuations, $\ln (1+\delta\rho/\rho)$.
The thermodynamic fluctuations can be in general described by a log-normal distribution (see example in Fig.~\ref{fig:PDF}), with small corrections due to high-order moments (skewness and kurtosis), thus 
strengthening the role of the power spectrum.
The log-normal distribution and weak non-Gaussian contributions are consistent with classic turbulence studies testing solenoidal stirring, albeit in non-stratified and controlled boxes (\citealt{Federrath:2010} and references therein), together with ICM studies (\citealt{Kawahara:2007,Zhuravleva:2013}).
Significant deviations start to arise with highly compressive turbulence, mainly affecting the wings (e.g.~\citealt{Kowal:2007}). We defer the study of
high-order moments to future work.

What can be inferred from X-ray observations? Besides X-ray imaging (see surface brightness maps in GC13, 
fig.~4, or \citealt{Churazov:2012}), X-ray energy spectra can provide crucial constraints on turbulence, in particular considering the forthcoming {\it Astro-H} mission (e.g.~\citealt{Inogamov:2003,Zhuravleva:2012,Tamura:2014}). We have two important tools to exploit, one is the broadening of the spectral line, and the other is the line shift.
In Fig.~\ref{fig:maps} (right), we show the
observed line broadening due to turbulent motions along the $y$-axis view. 
The Doppler broadening relative to the line rest energy $E_0$ is defined\footnote{The $\sqrt{2}$ term comes from the definition of the Gaussian distribution, $\propto \exp[-x^2/2\sigma^2]$, and not from isotropy arguments.} as
$\Delta E/E_0 \equiv \sqrt{2}\, \sigma_{\rm 1D, ew}/c$. The observed 1D velocity dispersion is computed as $\sigma^2_{\rm 
1D,ew}=E[v^2_{\rm 1D}]-E^2[v_{\rm 1D}]$, where $E[x]$ is the X-ray emission-weighted average along line of sight, using as emissivity $n_{\rm e}n_{\rm i}\Lambda(T)$ with X-ray threshold $T_{\rm x}\gta0.3$ keV (\citealt{Gaspari:2011a}).
The related full width of the line at half maximum is FWHM$\,\equiv2\sqrt{\ln 2}\,\Delta E\simeq1.66\,\Delta E$. The projected maps appear somewhat different from the cross-sections,
but the statistics, as the velocity dispersion, is the same after deprojection.
{\it Astro-H} will be able to resolve $\sim$4 eV (and {\it Athena} half this value); using the bright Fe XXV line at $6.7$ keV,
the lower detection limit becomes $\Delta E/E_0\sim0.06$ per cent (the dark regions in the right-hand panels of Fig.~\ref{fig:maps}). 
The maps show that, assuming good statistics, {\it Astro-H} could detect turbulence in most of the cluster, where $\sigma_{\rm 1D,ew}\gta200$ km s$^{-1}$ or 1D Mach number $\gta0.13$ (the non-black regions), covering our entire simulated range (see \citealt{Nagai:2013} for synthetic maps using {\it Astro-H} response).
Fe XXV is an excellent line since the associated thermal broadening is just $\sigma_{\rm th}= (k_{\rm b}T/56\,m_{\rm p})^{0.5}\sim120$ km s$^{-1}$, thereby the turbulent dispersion typically dominates the contribution to the total broadening of this line.

\begin{figure} %[!pth] %[!ht] %0.478
       \centering
       \subfigure{\includegraphics[scale=0.6]{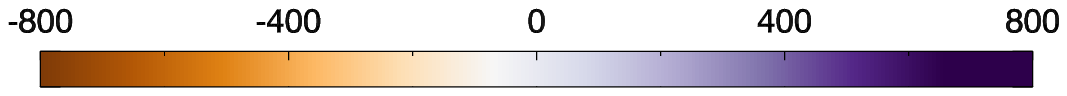}}
       \subfigure{\includegraphics[scale=0.6]{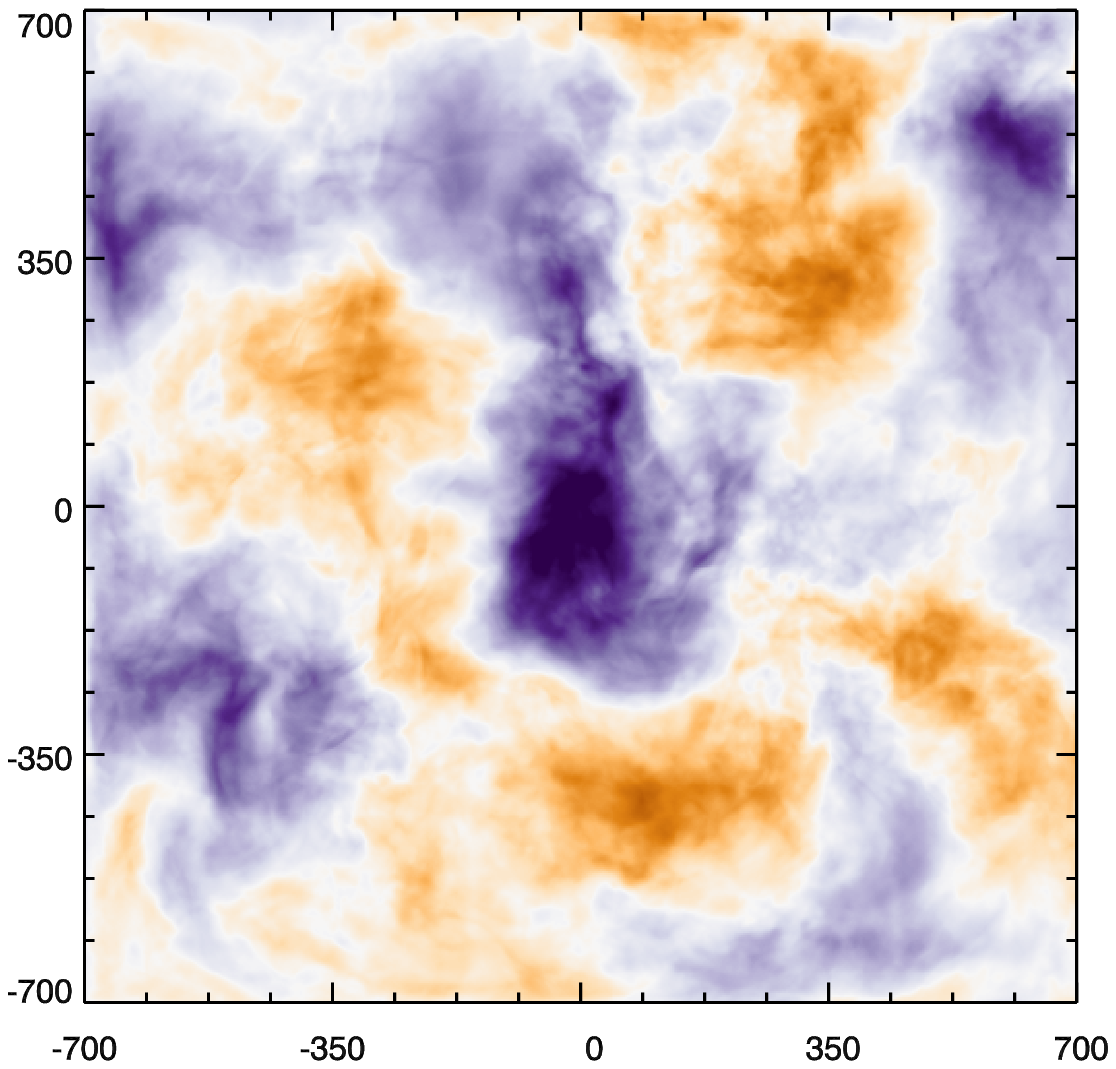}}                        
       \caption{X-ray emission-weighted velocity (km s$^{-1}$) along the $y$-axis, for the hydro model with $M\sim0.5$ (cf.~first row in Fig.~\ref{fig:maps}; the conductive models display similar maps). The projected velocity highlights only the large-scale motions, which dominate the kinetic energy content in our (and cosmological) runs, due to the injection at $L>100$ kpc. 
       }
     \label{fig:vemi}
\end{figure}

While the projected velocity dispersion $\sigma_{\rm 1D,ew}$ highlights the small-scale motions via the line broadening, the projected -- X-ray emission-weighted --  velocity field probes the large-scale motions via the line shift (e.g.~\citealt{Zhuravleva:2012}). Notice that the driven velocity field still has average 3D laminar motion $\sim0$.
Figure \ref{fig:vemi} shows the large eddies of size several 100 kpc, carrying most of the specific kinetic energy 
($\sigma_v^2/2\sim A_{v, \rm max}^2$), 
which would be absent in atmospheres stirred only at small scales.
The relative strength of the line broadening and shift thus carries important informations about the nature of the driven turbulence, a solid proxy corroborated by its insensitivity to conduction.

In passing, we note that X-ray observations are not the single channel to probe ICM fluctuations. 
We propose to use the thermal Sunyaev-Zel'dovich (SZ) effect to independently extract pressure fluctuations.
Current X-ray maps still have $\gta\,$$4\times$ higher spatial resolution. However, future observations (e.g. {\it ALMA, CCAT, SKA}) will allow to constrain the ICM power spectrum even in the submillimeter/radio band,
avoiding the use of expensive X-ray spectroscopy.

\section{Discussion and physical interpretation}  \label{s:disc} 
We discuss in this section the physical interpretation of the ICM power spectrum, in particular concerning the tight relation between the velocity and the other primary thermodynamic quantities (entropy, pressure, density). 
It shall be kept in mind that 
the reason why we perform 3D simulations is the impossibility to analytically solve a chaotic, nonlinear system. 
The following arguments arise from first order perturbations or dimensional theories, and shall be regarded as simple estimates.
The simulations generally confirm the ansatz presented below, albeit with relevant differences which are critically discussed.
We focus first on the normalization of the spectra ($l\sim L$; \S\ref{s:norm}), and then we analyze the spectral cascade ($l<L$; \S\ref{s:cascade}), along with the alterations imparted by conduction from the ideal evolution.

\subsection{Spectra normalization} \label{s:norm}
The normalization of the spectra is likely related to the relative importance of 
gravity waves and sound waves. Linearizing the perturbed hydrodynamic equations, it is possible to describe the propagation of a general wave in a spherical and gravitationally stratified atmosphere in terms of the following dispersion relation (see \citealt{Balbus:1990} for the WKBJ analysis):
\begin{equation}\label{e:disp}
\omega^4\;-\omega^2\,c^2_{\rm s}\, k^2 +\; \omega^2_{\rm BV}\, c^2_{\rm s}\, k^2_\perp  \,=\, 0, 
\end{equation}
where $k^2=k^2_r+k^2_\perp$ ($k_r$ and $k_\perp$ are the radial and azimuthal components of the wavenumber vector $\boldsymbol{k}$, respectively), $c_{\rm s}=(\gamma\, k_{\rm b}T/\mu m_{\rm p})^{1/2}$ is the adiabatic sound speed, and 
$\omega_{\rm BV}$ is the Brunt-V\"ais\"al\"a (buoyancy) frequency defined as
\begin{equation}\label{e:bv}
\omega_{\rm BV}\equiv\left[\frac{g}{\gamma}\,\frac{d\ln K}{dr}\right]^{1/2},
\end{equation}
where $g$ is the gravitational acceleration.
Eq.~\ref{e:disp} includes the action of two key waves. The middle term is associated with pressure waves, or simply sound waves ($p$-waves), while the last term represents gravity waves ($g$-waves) driven by the restoring buoyant force. 
In the next sections, we show that 
small perturbations driven by the two waves
are tied to the Mach number, $\sim \delta v/c_{\rm s}$.
For both $g$- and $p$-waves this holds within order unity (as in the simulated nonlinear regime), but the leading perturbations and dynamical modes differ: $g$-waves mainly drive entropy perturbations ($\delta K/K$), increase the gas vorticity, and induce a tangential bias in the turbulent velocity field;
$p$-waves are instead associated with compressive pressure fluctuations ($\delta P/P$),
a preferentially irrotational field, and isotropic turbulence (or with slightly radial bias).

\begin{figure} 
    \begin{center}
       \subfigure{\includegraphics[scale=0.38]{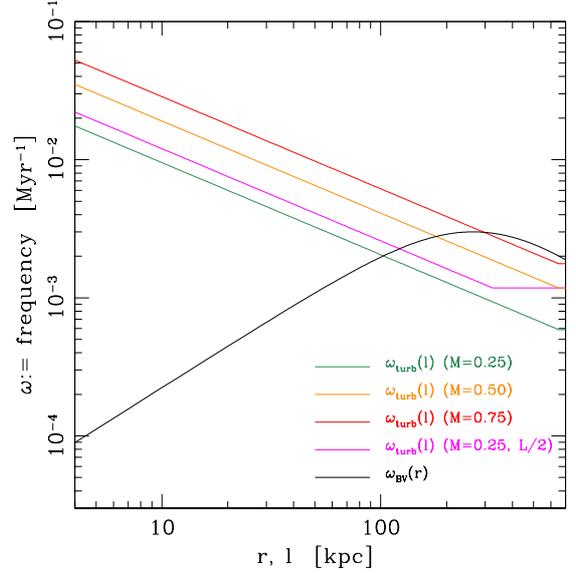}}
\caption{Typical frequency of the turbulent motions including the cascade (for the simulated sample $M=0.25 \rightarrow 0.75$), compared with the Brunt-V\"ais\"al\"a buoyancy frequency (black).
The minimum turbulence frequency is at the injection scale, typically $L\sim600$ kpc. For $\omega_{\rm turb} < \omega_{\rm BV}$ (Froude < 1), $g$-waves tend to be the process driving (entropy) fluctuations, while in the opposite regime sound waves drive stronger pressure perturbations, in both cases $\propto M$.
      }       
        \label{fig:freq}
     \end{center}
\end{figure}

In Figure \ref{fig:freq}, we show the frequency of the simulated turbulent motions, $t_{\rm turb}^{-1}$, compared with the Brunt-V\"ais\"al\"a frequency (black; Eq.~\ref{e:bv}) in the full radial range (at variance with GC13, where we focused on the properties of the central region). 
At large scales, the two frequencies tend to be roughly comparable (Froude number $\sim\,$1),
hence both $g$-waves and $p$-waves can be excited. This is a typical condition for most clusters, since large-scale profiles (entropy, pressure) are fairly self-similar, and turbulence follows our simulated subsonic range.
Let us first analyze the two limiting regimes, in order to understand better both processes. 

\subsubsection{Low frequencies (low $M$): $g$-waves} \label{s:gw}
The stratification of the ICM atmosphere allows to excite gravity waves (cf.~\citealt{Lufkin:1995,Ruszkowski:2010, Ruszkowski:2011}). In the low frequency regime, the dispersion relation in Eq.~\ref{e:disp} can be written as
\begin{equation}\label{e:lowf}
\omega^2 \simeq \omega^2_{\rm BV}\, \frac{k^2_\perp}{k^2},
\end{equation}
which tells us for $\omega > \omega_{\rm BV}$ gravity waves are evanescent, since $k_r$ must be imaginary. 
Therefore, wherever $\omega < \omega_{\rm BV}$ $g$-waves are excited.
For a cluster atmosphere, $\omega_{\rm BV}$ declines at large $r$ and waves are trapped within the radius such that 
$\omega\simeq\omega_{\rm BV}$ (\citealt{Balbus:1990}),
as $k\simeq k_\perp$.
More important, buoyant oscillations damp the radial component of turbulence, inducing a tangential bias in the gas velocity field (Froude < 1).
In the limiting case, the chaotic motions should collapse in azimuthal shells (e.g.~\citealt{Ruszkowski:2010}).
The profile of the velocity anisotropy parameter, $\beta \equiv 1-\sigma_{v_\perp}^2/2\sigma^2_{v_r}$ would show $\beta \ll 0$. Tangentially-biased vorticity is thus a good marker of the $g$-waves influence (see \S\ref{s:interp}).

Gravity waves are mainly tied to entropy perturbations
(see also Z14).
Using as dominant frequency $\omega_{\rm BV}$, the buoyant acceleration over 
a displacement $\delta r$ can be described with a simple harmonic oscillator:
\begin{equation}\label{e:aBV}
\ddot{r}_{\rm b} = -\, \omega_{\rm BV}^2\; \delta r = -\,\frac{c^2_{\rm s}}{\gamma^2\,|H_P|\,H_K}\;\delta r
\end{equation}
where $\omega_{\rm BV}$ is written\footnote{Assuming hydrostatic equilibrium, valid for low $M$, $g=-\,c^2_{\rm s}/(\gamma H_P)$; notice that $H_P<0$ and $H_K>0$.} as a function of both the scale height of entropy, $H_K\equiv dr/ d\ln K$, and pressure, $H_P\equiv dr/ d\ln P$.
Physically, $g$-waves occur because an entropy element is displaced from its equilibrium position, $r_0$, thus inducing an opposite force acting to restore the blob back to where the radial entropy is the same (clusters are convectively stable, $\boldsymbol{\nabla} K > 0$). The small displacement is thus linked to
$\delta K/K \simeq (d\ln K/dr)_0\, \delta r$, i.e.~$\delta r \simeq (\delta K/K)\, H_{K_0}$. 
The specific potential energy of the harmonic oscillator is $E_{\rm b}=\omega_{\rm BV}^2 (\delta r)^2/2$. Substituting for 
$\omega_{\rm BV}$ and $\delta r$, we can write
\begin{equation}\label{e:epot}
E_{\rm b}\simeq \frac{c_{\rm s}^2}{2\gamma^2}\frac{H_{K_0}}{|H_{P_0}|} \left(\frac{\delta K}{K}\right)^2.
\end{equation}
Using $E_{\rm b}$ as estimate for the average specific kinetic energy ($\bar{v}^2_{\rm 1D}/2\sim E_{\rm b}$), finally yields
\begin{equation}\label{e:relg}
\bar{M}_{\rm 1D} \sim \frac{1}{\gamma}\sqrt{\frac{H_{K_0}}{|H_{P_0}|}}\; \left|\frac{\delta K}{K}\right| \simeq  \mathcal{O}(1)\,\left|\frac{\delta K}{K}\right|.
\end{equation}
The last step arises from the fact that, for an isothermal atmosphere, the ratio of the scale heights is constant, $H_{K}/|H_{P}|=1/(\gamma-1)=1.5$. In general, $H_{K}/|H_{P}| = |\alpha_P|/\alpha_K\simeq 1-2$ (within $r_{500}$), where $\alpha_P$ and $\alpha_K$ are the slopes of the logarithmic radial profiles of pressure and entropy, respectively;
$\alpha_P$ steepens with increasing radius (\citealt{Arnaud:2010}), while $\alpha_K\simeq 1$.

\begin{figure} 
    \begin{center}
       \subfigure{\includegraphics[scale=0.52]{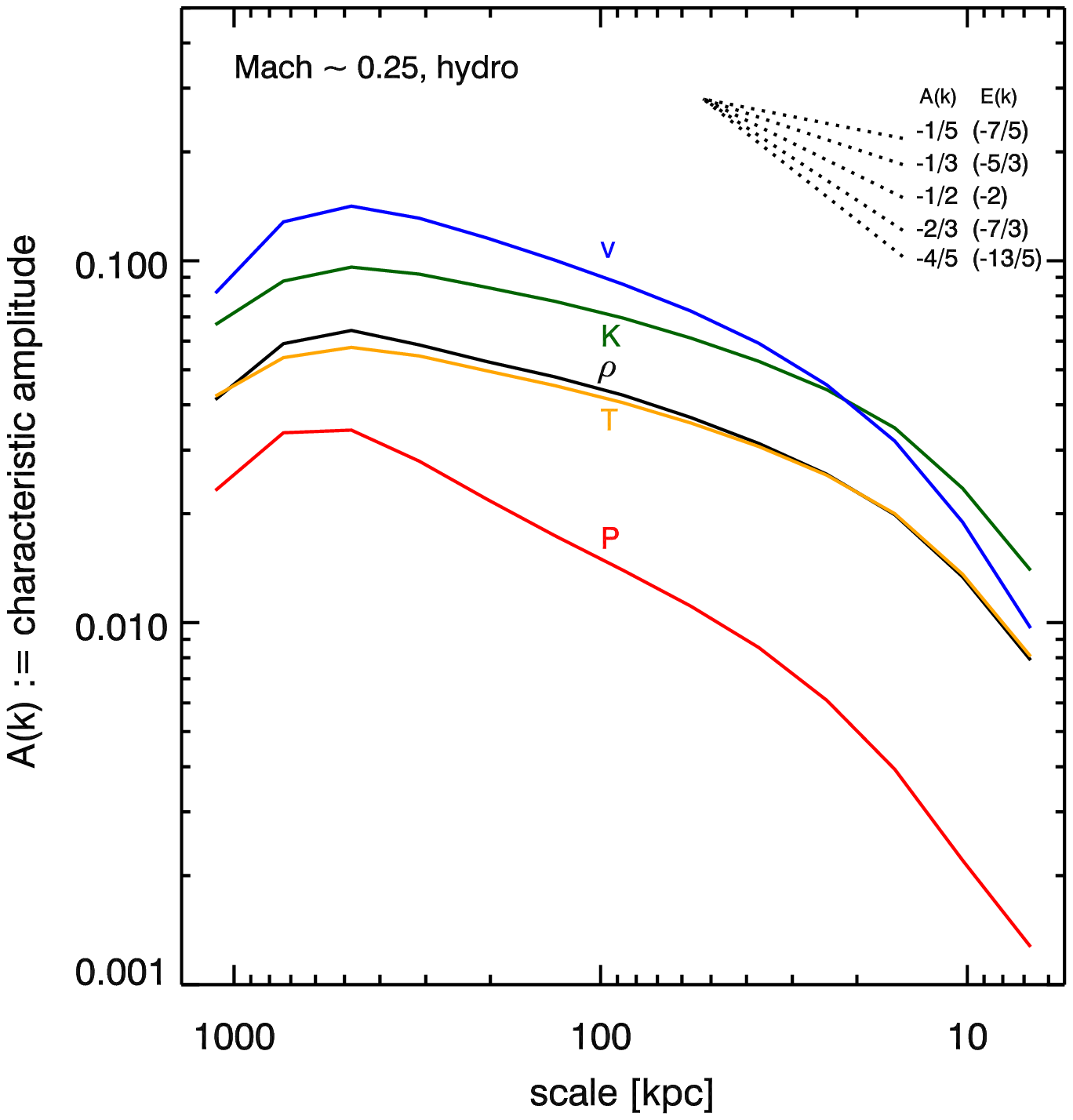}}
       \subfigure{\includegraphics[scale=0.52]{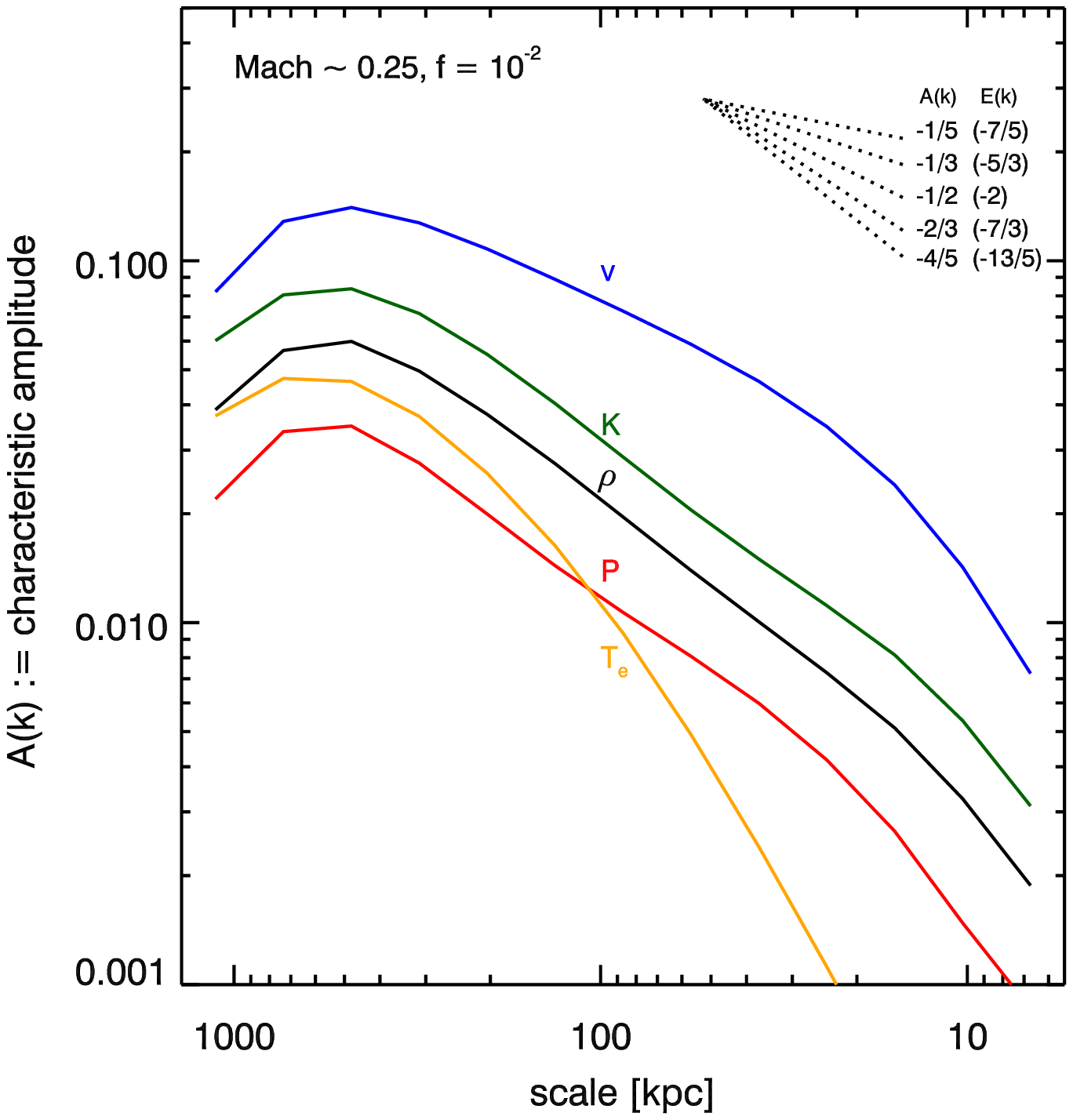}}       
\caption{Characteristic amplitude of the fluctuations related to all the thermodynamic quantities, for the $M\sim0.25$ flow without (top) and with conduction ($f=10^{-2}$; bottom): $v/c_{\rm s}$ (3D Mach), $\delta K/K$ (entropy), $\delta \rho/\rho$ (density), $\delta T/T$ (temperature), $\delta P/P$ (pressure). Except for turbulent velocities (laminar motions are null by construction), all other quantities are divided by the azimuthally averaged profile. For the conductive runs, we plot the (observable) electron temperature; the entropy parameter is $K\equiv(P_{\rm e}+P_{\rm i})/\rho^\gamma$. In the low $M$ flow, entropy perturbations (tied to $g$-waves) are the leading tracer of turbulent velocities, respecting the isobaric regime. Conduction gradually shifts the latter mode towards the isothermal regime, changing the relation with the derived thermodynamic quantities ($\delta K/K$ starts to approach density fluctuations).
      }       
        \label{fig:M025_all}
     \end{center}
\end{figure}  

Figure \ref{fig:M025_all} shows the power spectra of all the thermodynamic quantities, including turbulent velocities, for the hydro and a conductive run in the low Mach regime. Considering the hydro model (top),
the ratio of $M_{\rm 1D}\simeq M_{\rm 3D}/\sqrt{3}$ and $\delta K/K$ near the injection scale is 0.85, in line with the estimate in Eq.~\ref{e:relg}. Evidently, only one component of velocity is acting as efficient mixer, since cluster gradients are functions of $r$.
Fig.~\ref{fig:M025_all} clarifies that the low frequency regime corresponds to 
\begin{equation} \label{e:lowdp} 
\left|\frac{\delta P}{P}\right| \ll \left|\frac{\delta K}{K}\right|,
\end{equation}
since slow motions tend to be in pressure equilibrium with the surroundings, helped by the convective stability of the ICM.
The {\it isobaric} behavior is also manifest in the relation between density and temperature, or entropy and density  (both anticorrelated):
\begin{equation} \label{e:isob} 
\left|\frac{\delta \rho}{\rho}\right| \approx \left|\frac{\delta T}{T}\right|\ {\rm and}\ \left|\frac{\delta K}{K}\right| \approx \gamma\left| \frac{\delta \rho}{\rho}\right|\ \ \ \ \ {\rm [isobaric]}.
\end{equation}
Both relations are followed within $\lta\,$10 per cent, as shown in ~Fig.~\ref{fig:M025_all} (consistently with the Pearson analysis in GC13). Recall that halving the injection scale (\S\ref{s:spec}),
slightly reduces the strength of $g$-waves as $L^{1/3}$, since $\omega_{\rm turb}\propto L^{-1/3}$ approaches $\omega_{\rm BV}$ (Fig.~\ref{fig:freq}). $A_v/A_\rho$ thus increases by the same factor (sound waves are still too weak to contribute).

In the presence of mild conduction, the normalization is unaltered for $f< 10^{-2}$, although the intermediate cascade is damped by the increased diffusivity of the `tracer' (\S\ref{s:cascade}). For $f \gta10^{-2}$, also the large-scale perturbations are progressively driven towards the isothermal regime. Gravity waves can still induce a significant entropy contrast, but buoyancy is weakened by the increased diffusivity, tracing the shallower temperature gradient instead of
$\boldsymbol{\nabla} K$ (\citealt{Ruszkowski:2010}). The normalization of perturbations can thus decrease by a factor $\sim\,$2.
The relation between the different perturbations changes as
\begin{equation} \label{e:isoth} 
\left|\frac{\delta \rho}{\rho}\right| \gg \left|\frac{\delta T}{T}\right|\ {\rm and}\ \left|\frac{\delta K}{K}\right| \approx (\gamma-1)\left| \frac{\delta \rho}{\rho}\right|\ \ \ \ \ {\rm [isothermal]}.
\end{equation} 
Since turbulent regeneration is continuous and ions-electrons have a non-negligible equilibration time,
the pure isothermal regime is impossible to achieve. Nevertheless, as shown in Fig.~\ref{fig:M025_all} (bottom panel), the gap between entropy and density perturbations starts to shrink, as $T_{\rm e}$ fluctuations gradually lower (see also Fig.~\ref{fig:Mhigh_all}, bottom).

\subsubsection{High frequency (high $M$): $p$-waves} \label{s:pw}

In the opposite regime, i.e.~high frequency ($M>0.5$), the dispersion relation (Eq.~\ref{e:disp}) is shaped by the contribution of sound waves ($p$-waves), which can be now written as
\begin{equation}\label{e:highf}
\omega^2 \simeq c_{\rm s}^2\, k^2.
\end{equation}
The azimuthal component of the wavenumber scales as $k_\perp\propto r^{-1}$. Therefore, $p$-waves excited in the cluster central regions become mainly radial further out, $k\sim k_r$.
Moreover, disturbances to the vorticity results to be proportional to (\citealt{Lufkin:1995})
\begin{equation}\label{e:irrot}
\frac{\partial}{\partial t} \delta \left(\boldsymbol{\nabla} \times \boldsymbol{v}\right) \propto \boldsymbol{k}\times \boldsymbol{\nabla} \ln K,
\end{equation}
implying that $p$-waves are preferentially\footnote{Vorticity can be in part generated via the baroclinic instability, i.e.~as sound waves travel obliquely across the entropy gradient.} 
characterized by an irrotational velocity field, in contrast with the more
tangential $g$-waves. Overall, stronger $p$-waves tend to restore isotropic turbulence, or to induce a slightly radial bias (depending on how many sound waves are excited in the central regions; \S\ref{s:interp}).

\begin{figure} 
    \begin{center}
       \subfigure{\includegraphics[scale=0.505]{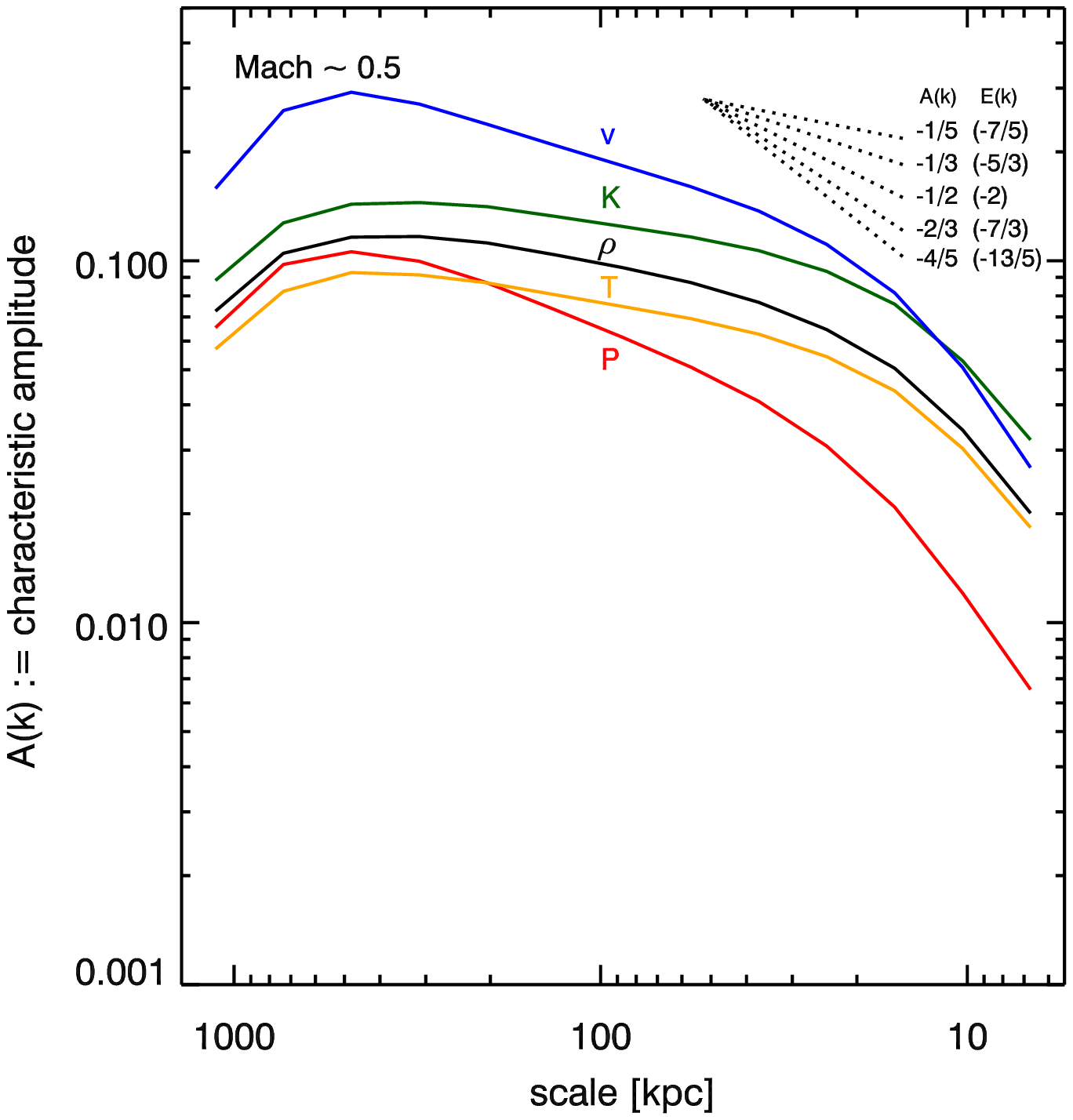}}
       \subfigure{\includegraphics[scale=0.505]{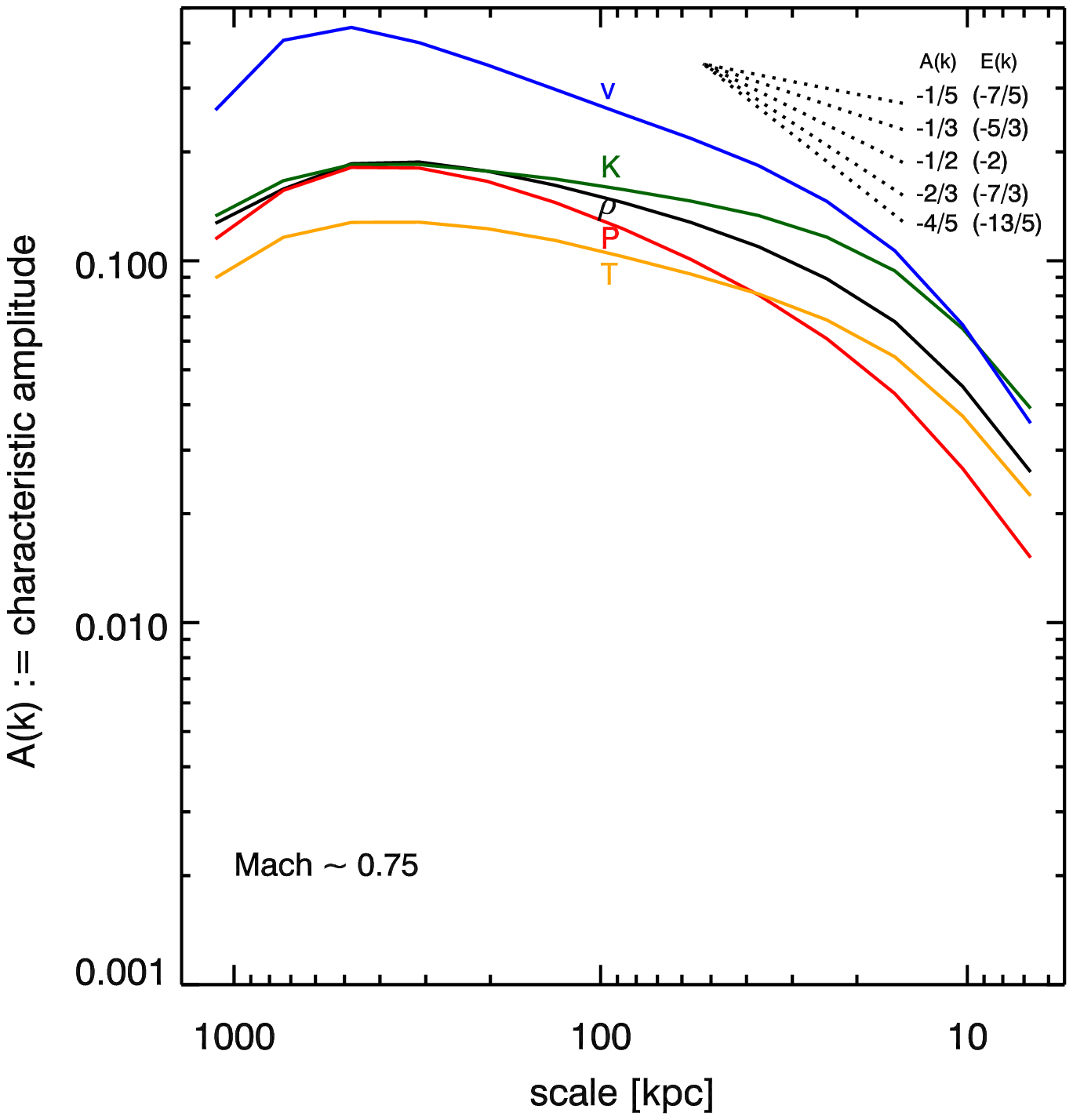}}
       \subfigure{\includegraphics[scale=0.505]{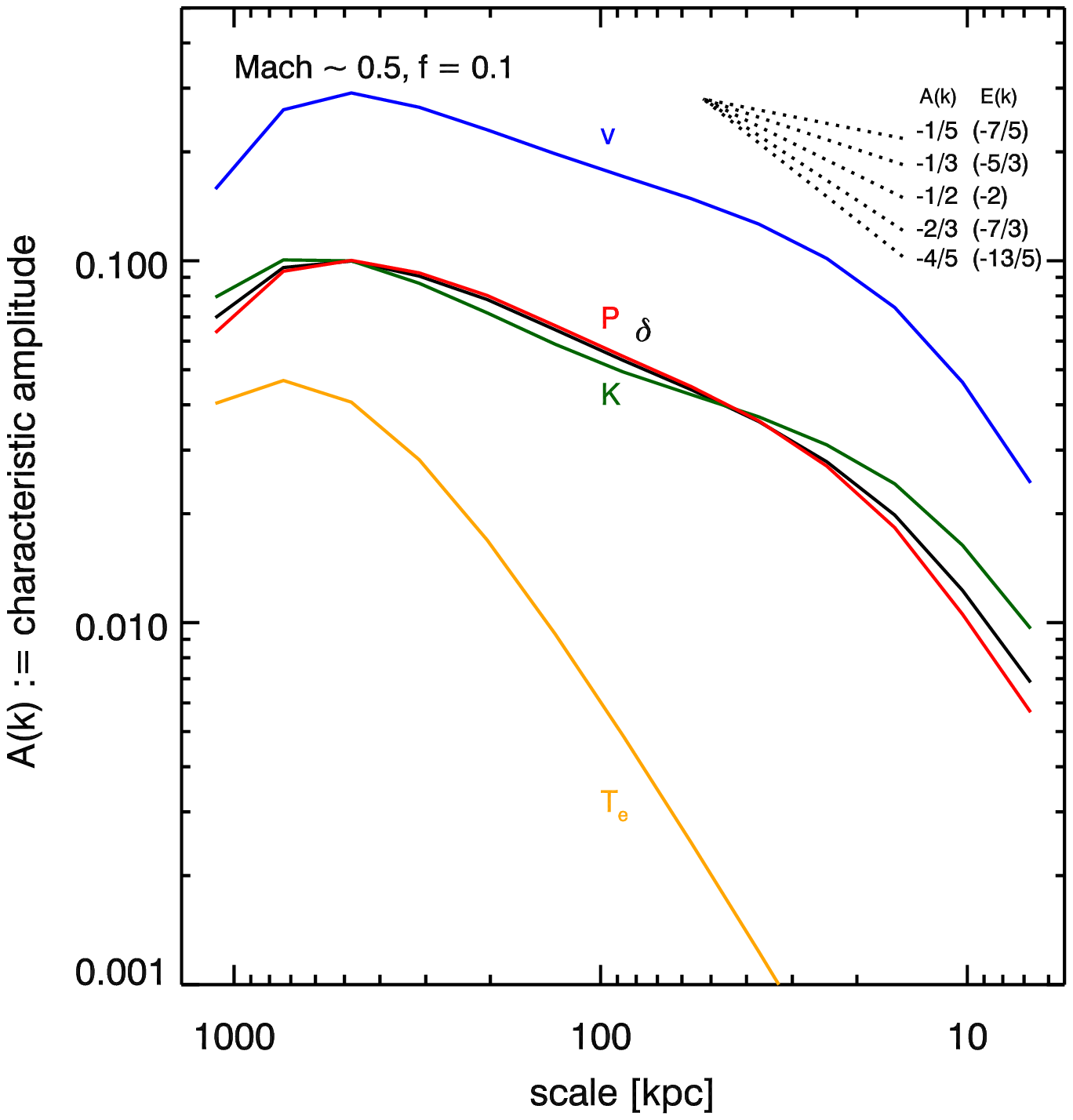}}                     
\caption{Characteristic amplitude of all the thermodynamic fluctuations, for the runs with $M\sim0.5$ (hydro and $f=0.1$) and $M\sim0.75$ (cf. Fig.~\ref{fig:M025_all}). In significantly turbulent atmospheres, $p$-waves start to affect the fluctuations dynamics via $\delta P/P$, still in conjunction with entropy perturbations ($\omega_{\rm turb}$ is not yet $\gg\omega_{\rm BV}$). The derived quantities follow from the adiabatic/isothermal mode, for the hydro/conductive flow.
      }       
        \label{fig:Mhigh_all}
     \end{center}
\end{figure}

The characteristic injection frequency of $p$-waves is
$\omega^2_s\simeq c_{\rm s}^2/ L^2$. In this high frequency regime
\begin{equation} \label{e:highdp} 
\left|\frac{\delta P}{P}\right| > \left|\frac{\delta K}{K}\right|,
\end{equation}
meaning that pressure perturbations drive the dynamics and fluctuations follow the {\it adiabatic} regime (constant entropy). Following the same arguments provided in the previous section, the displacement magnitude is now tied to pressure variations as $|(\delta P/P)\,H_{P_0}|$. Using the potential energy as estimate for the average kinetic energy (cf.~Eq.~\ref{e:epot}-\ref{e:relg}) now yields
\begin{equation}\label{e:relp}
\bar{M}_{\rm 1D} \sim \frac{|H_{P_0}|}{L} \left|\frac{\delta P}{P}\right| \simeq  \mathcal{O}(1)\,\left|\frac{\delta P}{P}\right|,
\end{equation}
where $|H_{P_0}|\simeq L/|\alpha_P|$, since we are analyzing the large-scale power.
We note that stratification allows a more efficient generation of sound waves compared with uniform media (e.g.~\citealt{Stein:1967}), as shown by the simulations, due to the partial conversion of solenoidal turbulence in more compressive motions.

Comparing the last estimate with Eq.~\ref{e:relg}, it is clear that in both cases the (1D) Mach number drives the spectrum normalization of perturbations, as found in the simulations. The transition must be smooth, as hinted by the general dispersion relation (Eq.~\ref{e:disp}).
However, the driving perturbations change character. Figure \ref{fig:Mhigh_all} shows that for $M\gta0.5$, $\delta P/P$ rises linearly with $M$, while $\delta K/K$ remains constant ($A_{K, {\rm max}}\sim0.15$ in both hydro runs). In the $M\sim0.75$ flow (middle panel), $\delta P/P$ has reached $\delta K/K$, hence $p$-waves do not yet fully overcome $g$-waves, which would happen for $M\gta1$ (as $\omega_{\rm turb} > \omega_{\rm BV}$ at each radius, or Froude $> 1$; Fig.~\ref{fig:freq}).
Moreover, while gravity waves tend to accumulate in the system, $p$-waves may leave it in a few sound-crossing times.
Again, turbulence with high Mach number is required to see a system fully dominated by $p$-waves.

According to Eq.~\ref{e:highdp}, the perturbations of the other thermodynamic quantities start to shift from the isobaric to adiabatic regime, implying the following conversion:
\begin{equation} \label{e:adia} 
\left|\frac{\delta \rho}{\rho}\right| \approx \frac{1}{\gamma-1}\left|\frac{\delta T}{T}\right|\ {\rm and}\ \left|\frac{\delta P}{P}\right| \approx \gamma\left| \frac{\delta \rho}{\rho}\right|\ \ \ \ \ {\rm [adiabatic]},
\end{equation}
as signaled by the increasing gap between density and temperature fluctuations (up to $\simeq1.5$).
Adding conduction (Fig.~\ref{fig:Mhigh_all}, bottom) shifts again the adiabatic mode towards a partial isothermal regime (Eq.~\ref{e:isoth}).
Interestingly, pressure fluctuations are now more clearly the driver of density fluctuations ($\delta P/P \sim \delta\rho/\rho$), since $\delta K/K$ has degraded by over 50\% compared with the hydro run.  

We note that, in the low $M$ regime, the presence of weak $\delta P/P$ can be mainly attributed to
the conservation of the Bernoulli parameter (although exactly valid only in the steady state). 
For a constant potential, it can be written as
\begin{equation} \label{e:Bern} 
\frac{v^2}{2} + \frac{c^2_{\rm s}}{\gamma-1}={\rm const.}
\end{equation}
Differentiating and taking as reference velocity $v_0 = 0$, yields $\delta P/P\propto M^2$,
which is indeed followed by the runs with $M< 0.5$ (compare Fig.~\ref{fig:M025_all} and \ref{fig:Mhigh_all}, top panels), while models with stronger turbulence follow $\delta P/P\propto M$ (Fig.~\ref{fig:Mhigh_all}, top and middle panel).

\subsubsection{Real systems: $g$- and $p$-waves interplay} \label{s:interp}

\begin{figure} 
    \begin{center}
       \subfigure{\includegraphics[scale=0.36]{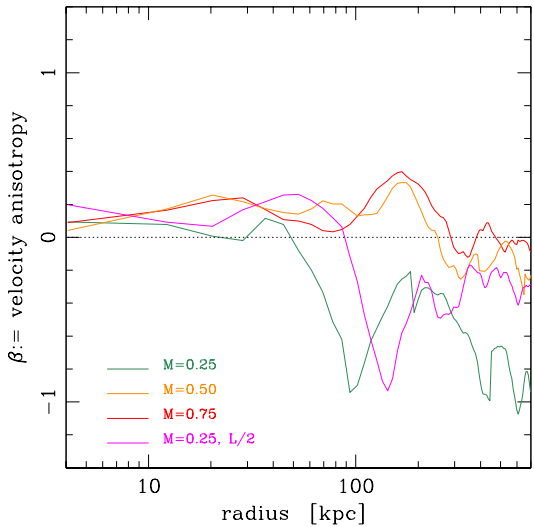}}
\caption{Velocity anisotropy as a function of radius, $\beta(r) \equiv 1-\sigma_{v_\perp}^2/2\sigma^2_{v_r}$, for all the hydro models. Negative/positive values imply a tangential/radial bias. $g$-waves ($M\le 0.25$) tend to damp the radial component of turbulent motions, inducing a tangential bias; $p$-waves ($M\ge0.75$) tend instead to preserve isotropic motions, or to induce a slightly radial bias. In realistic clusters, the anisotropy is expected to be minor, due to the interplay of both waves and the recurrent stirring.
      }       
        \label{fig:beta}
     \end{center}
\end{figure}  

The estimates in Eq.~\ref{e:relg} and \ref{e:relp} are crude approximations to reality. In the realistic cluster evolution, the 3D hydrodynamic equations, and the related perturbations, are nonlinear, with the addition of chaotic stirring (\S\ref{s:dep} for other deviations). Multiple frequencies act at the same time, describing waves at different scales and radii. Although the simulations confirm that the spectra normalization is provided by $M_{\rm 1D}$, we expect a combination of $g$- and $p$-waves shaping the dynamics of a turbulent cluster. This is visually highlighted by the maps in Fig.~\ref{fig:maps}. More quantitatively, we can discriminate the action of both channels through an important marker, i.e.~the anisotropy of velocities along the radial direction (\S\ref{s:gw}-\ref{s:pw}). Notice that computing the velocity spectra over the scale $l$ obfuscates the radial anisotropy.

In Figure \ref{fig:beta}, we show the $\beta \equiv 1-\sigma_{v_\perp}^2/2\sigma^2_{v_r}$ parameter for all the hydro models. For the low $M\sim0.25$ flow, above $\sim$100 kpc the motions are mildly\footnote{The retrieved anisotropy is not strong, since $\beta\simeq-1$ corresponds to $v_\phi \simeq \sqrt{1-\beta}\,v_r \simeq 1.4\,v_r$, difficult to observe by visual inspection.} tangential, reaching a minimum value $\beta\sim-1$. 
Analyzing the frequencies in Fig.~\ref{fig:freq}, this corresponds to the region where $\omega < \omega_{\rm BV}$, i.e.~where $g$-waves tend to be more dominant, damping the radial component of turbulent motions.
Raising $M$ increases the turbulence frequency up to $\sim\omega_{\rm BV}$, even near the injection scale at several 100s kpc. Therefore, the large-scale motions progressively lose the tangential bias, drifting towards isotropy ($\beta \sim 0$). In the $M\sim0.75$ flow (red), $p$-waves start to dominate: motions show no sign of the tangential bias, with instead a slightly radial bias (\S\ref{s:pw}).
Within $r<100$ kpc, the frequencies of the turbulent cascade are always greater than $\omega_{\rm BV}$, hence $p$-waves start to have a major influence. The transition occurs at smaller radii for lower $M$, as suggested by Fig.~\ref{fig:beta}.  
On the other hand, large-scale $g$-waves tend to be trapped within the cluster core; the combination of the two effects results in a quasi isotropic $\beta$ in the core (or slightly radial $\beta \sim0.2$). In the presence of conduction, the same scenario applies, but the $\beta$ factor is globally reduced (by $\sim$2), towards the isotropic value. As noted in \S\ref{s:gw}, thermal conduction indeed inhibits buoyancy.

Overall, we suggest to use the $\beta$ parameter to discriminate the effects of $g$- and $p$-waves, in conjunction with the thermodynamic mode. We remark that typical cluster conditions are expected to show at best mild anisotropic motions, and emerging only in spherical coordinates; most clusters are not in a regime in which $\omega\ll\omega_{\rm BV}$ (or the opposite), where radial motions are dramatically suppressed. The recurrent stirring also promotes $\beta$ values drifting towards isotropy.
Nevertheless, the ICM spectrum normalization is always comparable to $M_{\rm 1D}$ within order unity, 
regardless of which one is the driving wave, and for both the linear approximation (Eq.~\ref{e:disp}) and the nonlinear simulations.

\subsection{Spectral cascade: advection of tracers} \label{s:cascade}
We have analyzed so far the physical interpretation of the spectra normalization ($l\sim L$). The next question is why do perturbations show an inertial cascade similar to that of velocities ($l<L$)? 
The main thermodynamic variables can be crudely considered as `tracers' of the velocity field, whose spectra are explained with the advection theory of passive scalars (e.g.~\citealt{Sreenivasan:1991,Monin:1975,Warhaft:2000} for a review). This is particularly relevant in the subsonic regime, since the compressive term $\boldsymbol{\nabla}\cdot \boldsymbol{v}\rightarrow 0$. 

The equation governing the advection of a passive incompressible scalar $C$ is given by (\citealt{Warhaft:2000}, sec.~1)
\begin{equation}\label{e:tracer}
\frac{DC}{Dt}\equiv \frac{\partial C}{\partial t} + {\boldsymbol v}\cdot{\boldsymbol\nabla}C = \boldsymbol{\nabla}\cdot (\kappa\, \boldsymbol{\nabla} C),
\end{equation}
where $\kappa$ is the diffusivity of the tracer and $D/Dt$ is the Lagrangian derivative. 
According to the classic Kolmogorov-Obukhov-Corrsin theory (KOC; \citealt{Obukhov:1949,Corrsin:1951}; \citealt{Warhaft:2000}), the energy spectrum of the scalar linearly traces that of velocities, i.e.
\begin{equation}\label{e:KOC}
E_C(k)\propto E_v(k)\propto k^{-5/3}.
\end{equation}
Considering a 2D vortical motion, the vorticity is conserved (see also Kelvin's theorem), 
$D(\boldsymbol{\nabla} \times \boldsymbol{v})/Dt=0$, in analogy to Eq.~\ref{e:tracer}. The turbulent eddies and the scalar are thus expected to share similar properties, like the spectral cascade, although the diffusive term will introduce some discrepancies (\S\ref{s:dep}).

Entropy is an excellent example of `passive tracer'.
$S \equiv  k_{\rm B}/[(\gamma-1)\mu m_{\rm p}]  \, \ln K$
has indeed the advantage of being insensitive to adiabatic compressions or expansions.
Aside diffusion, the lagrangian derivative of entropy is thus conserved,  if no irreversible heating ($\mathcal{H}$) or cooling ($\mathcal{L}$) occurs, such as
\begin{equation}\label{e:entropy}
\rho T \frac{DS}{Dt}\equiv\rho T \left(\frac{\partial S}{\partial t} + {\boldsymbol v}\cdot{\boldsymbol\nabla}S\right) = \mathcal{H}-\mathcal{L}\simeq\boldsymbol{\nabla}\cdot (D_{\rm turb}\, \rho T\,\boldsymbol{\nabla} S),
\end{equation}

The hot ICM has negligible radiative cooling ($\mathcal{L}\simeq0$);
the only source of heating can be turbulent diffusion or dissipation. The latter is subdominant for subsonic flows, $t_{\rm diss, heat}\simeq M^{-2}\,t_{\rm turb}$.
The entropy $S$ can thus replace the scalar $C$ in Eq.~\ref{e:tracer}, with a diffusivity tied to the turbulent field, $D_{\rm turb}\sim\sigma_{v}\,l$ (\S\ref{s:init}). 
In the low $M$ regime (\S\ref{s:gw}), entropy fluctuations tend to lead the dynamics of perturbations, linked to the large-scale $g$-waves. The $\delta K/K$ cascade then develops over $l<L$, tracing the velocity inertial regime (Fig.~\ref{fig:M025_all}), in line with KOC theory (Eq.~\ref{e:KOC}).
Since the fluctuations of density and temperature follow from the dominant mode -- isobaric (Eq.~\ref{e:isob}), isothermal (Eq.~\ref{e:isoth}), or adiabatic (Eq.~\ref{e:adia}) --, also their cascade traces that of $\delta K/K$. 
As discussed before, increasing the Mach number, boosts the impact of $p$-waves. The leading tracer gradually shifts towards $\delta P/P$ fluctuations. In the $M\sim0.75$ hydro run (Fig.~\ref{fig:Mhigh_all}), density perturbations start to track more closely the $\delta P/P$ cascade. Pressure is affected by adiabatic processes, yet the simulations show that its cascade is slightly steeper than that of velocities, signaling that it can be used as a crude tracer of the (subsonic) flow. Interestingly, pressure is not directly affected by turbulent mixing (which acts on entropy), and thus displays a tighter cascade with velocities\footnote{In the $M \sim 0.25$ run, the $\delta P/P$ cascade is instead steeper, following the classical $E_P\propto k^{-7/3}$, solely driven by the Bernoulli term (Eq.~\ref{e:Bern}). }. 

Adding conduction, increases the effective diffusivity of the tracer ($\kappa \rightarrow f \kappa_{\rm S}$), leading to a cascade steeper than that of Kolmogorov (Fig.~\ref{fig:M025_all}, bottom; notice how $T_{\rm e}$ fluctuations are strongly damped).
The decline of the tracer spectrum starts to occur as $D_{\rm cond}> D_{\rm turb}/100$, i.e.~$P_{\rm t} < 100$. 
Physically, the quick increase in temperature induces a rapid re-expansion of the forming overdensity, overcoming the 
action of turbulent regeneration.
In the conductive regime, the $A_v/A_\rho$ ratio is thus expected to gradually increase as a function of $f$ (Fig.~\ref{fig:ratio}).
The ratio widens up to a factor of 3$\,$-$\,$5 over all scales in the presence of strong conduction ($f\gta0.1$). 

It is interesting to note that 
in the simple KOC picture
\begin{equation}\label{e:Enorm}
E(k)_C = b\: \bar{\epsilon}_C^{\,-1/3}\, \bar{\epsilon}_v \,\, k^{-5/3},
\end{equation}
where $b$ is a universal constant.
The average dissipation rate of the passive tracer and velocity is 
$\bar{\epsilon}_C\sim \sigma_C^2/(L/\sigma_v)$ and 
$\bar{\epsilon}_v \sim \sigma_v^3/L$,  
respectively, implying that the normalization of Eq.~\ref{e:Enorm} is {\it independent} of $\sigma_v$. 
In reality, the compressive term sustains a relation between the density variance and the Mach number, even in homogeneous media: in the subsonic range $\delta\rho/\rho$ fade as $M^2$ (cf.~\S\ref{s:pw}), while the relation becomes linear for supersonic turbulence (see \citealt{Kowal:2007} and \S\ref{s:intro}).
The latter is however conceptually different from our retrieved linear relation developing in the subsonic state of galaxy clusters. In the {\it stratified} ICM plasma,
increasing Mach number implies larger coherent displacement, leading to a larger contrast of entropy/pressure defined by the cluster gradients or scale heights (Eq.~\ref{e:relg} and \ref{e:relp}).

\subsubsection{KOC departures and the radial gradients}\label{s:dep}
Although KOC theory can explain the global picture of the spectral cascade, its arguments are purely based on dimensional analysis.
Our retrieved slope of entropy/density is typically shallower than the Kolmogorov index, in the non-diffusive models.
Physical experiments (e.g.~fig.~5 in \citealt{Sreenivasan:1991}) show that the tracer slope approaches the Kolmogorov cascade only for very high Reynolds numbers and in a slow asymptotic way. For low Reynolds numbers, as in our ICM simulations ($R_L\lta500$; GC13, sec.~2.6), the spectral index is expected to be shallow (cf.~figure 4 in \citealt{Warhaft:2000}). 
Even pressure fluctuations, albeit in line with the Kolmogorov cascade, are shallower than the classic expectation $E_P\propto k^{-7/3}$ ($A_P\propto k^{-2/3}$; \citealt{Schuecker:2004}), in the $M\gta0.5$ runs.
A slope as shallow as $A_k\propto k^{-1/5}$ signals that the timescale for transferring the tracer variance from large to small scales is $\propto k^{-4/5}$, instead of the Kolmogorov $\propto k^{-2/3}$,
due to diffusion effects affecting the transfer process in different ways (see the viscosity test in Fig.~\ref{fig:core}).
This is remarked
by the uncorrelated phases between velocity and the tracer (Fig.~\ref{fig:maps}). 

Another departure from the KOC cascade may be associated to compressive features.
In the extreme case of highly supersonic turbulence, shocks induce very thin peaks in gas density
(\citealt{Kim:2005}, fig.~2). Sharp peaks can be seen as delta functions, which in Fourier space generate a flat spectrum, $P_\delta\sim k^0$. ICM turbulence is however subsonic, thus the contribution of thin compressive features to our observed flattening is limited.

The incomplete similarity with the Kolmogorov cascade more likely depends on the initial entropy/pressure gradients.
In fact, Eq.~\ref{e:relg} and \ref{e:relp} are only valid for small displacements. For a nonlinear evolution, the injected eddy will experience $H_K$ and $H_P$ varying with radius, given that the initial cluster profiles are self-similar power laws (\S\ref{s:gw}):
\begin{equation}\label{e:igrad}
\frac{\delta K}{K} = \alpha_K\, \frac{\delta r}{r}\ \ {\rm and}\ \ \frac{\delta P}{P} = \alpha_P\, \frac{\delta r}{r}.
\end{equation}
Fixing for instance $\delta r\sim L$,
the injected turbulence at smaller radii can create relatively larger contrasts, inducing a flattening in the spectral cascade.
The magnitude of this effect depends however on the relation $r \longleftrightarrow l$. Since chaotic motions are 3D, the turbulent streamlines intersect different projections of the radial gradients,
hence the dependence shall be weak.
Neglecting the previous KOC departures,
the flattening of the simulated spectra corresponds to an average $r\sim l^{0.13}$. 
An opposing effect is related to the fact that
the indices $\alpha_K$ and $\alpha_P$ are not constant, but both declines within the core radius (turbulent mixing also slightly lowers $\alpha_K$ in time; GC13).
According to Eq.~\ref{e:igrad}, a lower slope implies a lower contrast.
This may explain why the $A_v/A_\rho$ ratio slowly declines towards smaller scales (Fig.~\ref{fig:ratio}, top), 
though always remaining larger than unity. We will investigate in future other cluster atmospheres, to assess the impact of different $K$ and $P$ scale heights.  
In closing, we note that all these secondary effects are washed out in the presence of any significant diffusivity, which completely inverts the $A_v/A_\rho$ downtrend (Fig.~\ref{fig:ratio}).

\subsection{Further improvements}\label{s:imp}
Finally, we discuss the limitations of the models and further improvements.
We studied here the evolution of the intracluster plasma, primarily in the hot regime.
In future works, we plan to extend the simulated sample (e.g.~strong cool-core systems) and to test additional physics.
Needless to say, 3D high-resolution 2T simulations with turbulence and diffusive terms are extremely expensive, hence small steps must be taken.

It will be interesting to include the effect of cooling, which can induce thermal instability and condensation of cold filaments (\citealt{Gaspari:2012a,Gaspari:2013_cca}). AGN feedback balances cooling, preserving the cluster core in global quasi-thermal equilibrium  
(e.g.~\citealt{Gaspari:2012b}).
However, both processes just affect the inner region $r<0.1\,R_{500}$ (cf.~\citealt{Gaspari:2014}), while at large radii galaxy clusters maintain self-similarity, especially in the entropy profile ($\alpha_K\sim1$; e.g.~\citealt{Panagoulia:2014}). We thus do not expect dramatic deviations from the current ICM power spectrum (which is intrinsically volume-weighted) and we believe our results can be applied to a wide range of clusters and conditions.
Strongly unrelaxed systems, as major mergers, might present significant variations, e.g.~due to the dynamic gravitational potential, and requires to be further tested. 
We are also studying the role of very small injection scales (e.g.~AGN outflows): for $L< 50$ kpc entropy perturbations may be considerably weaker (as $\omega_{\rm BV}< \omega_{\rm turb}$), while pressure perturbations should drive the normalization of the density power spectrum, even at low Mach numbers.

In the companion work (Z14), we improve the driving, including the cosmological evolution and the turbulence generated by mergers and large-scale inflows. Albeit limited by low resolution, we find that the linear $M_{\rm 1D}$-$\delta\rho/\rho$ relation holds across a large sample of simulated clusters. 
We retrieve a relation scatter of $\sim\,$30 per cent. In the cosmological context, it is more difficult to disentangle the source of the velocity anisotropy, especially in unrelaxed systems.
As for observational data, it is important to accurately remove the underlying radial profile and the strongly nonlinear sub-structures, which can contaminate the large-scale power. We find that the most reliable scales dominated by the turbulent cascade are $l<300$ kpc, which is fortunately the optimal regime for X-ray observations (see GC13, sec.~4.3 for a comparison with real data).

We plan to test additional physics. We currently probed the effects of (nearly maximal) Spitzer-like viscosity and electron-ion equilibration. It will be interesting to assess the role of the related magnetic suppression factors 
(which can be different from that of heat transport). 
On the other hand, diffusivities linked to the ions are roughly two orders of magnitude slower compared with electron thermal conduction, since the electron sound speed is $\simeq\,$43 times that of ions. We thus expect conduction to dominate the shape of the power spectrum over a large range of scales.
A viscosity lower than the present Spitzer-like value (which would require a much higher resolution) would imply a more extended inertial cascade. Compared with Fig.~\ref{fig:ratio}, $A_v/A_\rho$ should thus differ only below 10s kpc, continuing to widen in combination with high conductivity. In the presence instead of both low viscosity and conductivity, density and velocity spectra are expected to be tightly coupled again (KOC theory).

Fully MHD simulations are a further route of improvement, modeling better the local features (as cold fronts and filaments). However, besides the numerical complication of integrating anisotropic conduction for long times, MHD runs can at best retrieve the geometric suppression factor (see \S\ref{s:init}). Therefore, we would still be forced to parametrize the conductivity with a factor $f_\parallel$, in order to include microinstabilities and line divergence below the gas mean free path.

\section{Conclusions}\label{s:conc}
We carried out 3D high-resolution hydrodynamic  2T simulations, in order to study the power spectrum of the hot intracluster plasma in its various manifestations. 
We focused on the properties of the velocity field and the intimate relation with the driven thermodynamic fluctuations (in particular of density, the primary observable). The ICM power spectrum contains enough information to accurately constrain the dominant physics of the diffuse medium, as the strength of turbulent motions, the level of thermal diffusivity, and the thermodynamic mode, among the most notable.
The spectra of $v/c_{\rm s}$ and of perturbations (e.g.~$\delta\rho/\rho$) are globally self-similar, varying the strength of turbulence via the 3D Mach number, $M$, or changing the injection scale, $L$. \\
At the large cluster scales ($l\sim L$), i.e.~several 100 kpc:
\begin{itemize}
\item
Weak turbulent motions in the cluster ($M\lta0.25$) mainly excite gravity waves ($\omega_{\rm turb}<\omega_{\rm BV}$); the leading perturbations are related to entropy variations $\delta K/K$.
For stronger turbulence ($M > 0.5$), sound waves start to significantly contribute ($\omega\gta\omega_{\rm BV}$), 
passing the leading role to the compressive pressure fluctuations $\delta P/P$.\\

\item
The other thermodynamic perturbations, as $\delta \rho/\rho$ and $\delta T/T$, derive from the dominant mode of the process: isobaric (for $g$-waves/low $M$), adiabatic (for $p-$waves/high $M$), or a mixed state for intermediate $M$.
Conduction shifts the perturbations towards the isothermal mode.\\

\item
In both the regimes driven by $g$- or $p$-waves, the turbulent 1D Mach number is comparable to the variance of the leading perturbations ($K$ or $P$), within order unity. E.g.~for $M\lta0.25$ flows, $M_{\rm 1D} \sim \delta K/K \sim \gamma \,\delta \rho/\rho$. 
Quantitatively, all simulations show a {\it linear relation} given by
$A_{v,\rm max}\,$$\simeq\,$$2.3 \,A_{\rho,\rm max}$ (at $L$$\,\sim\,$600 kpc), with a weak $L^{1/3}$ scaling. We remark that to convert between Fourier and real space the relation to apply is instead $M\approx 4\,{A_{\rho,\rm max}}$ (at $L$$\,\sim\,$600 kpc).
\\

\item
Turbulent motions with a tangential bias ($\beta(r)<0$) mark the influence of $g$-waves (low $M$), 
while $p$-waves (high $M$) tend to preserve isotropy or to induce a slightly radial bias.  
Most clusters show intermediate Mach numbers,
hence we expect a mixed regime drifting towards global isotropy  (Froude $\sim\,$1).
\\

\end{itemize}

\noindent
At the intermediate/small scales ($l < L$), i.e.~10$\,$-$\,$100 kpc:
\begin{itemize}
\item
The turbulent velocities develop a Kolmogorov cascade ($A_v\propto k^{-1/3}$ or $E_v\propto k^{-5/3}$) in all subsonic runs, despite stratification ($\omega_{\rm turb} < \omega_{\rm BV}$). The thermodynamic perturbations, in particular entropy, act as effective `tracers' of the velocity field, developing an analogous inertial cascade, in line with the classic (Kolmogorov-Obukhov-Corrsin) advection theory of passive scalars in turbulent media.\\

\item
The cluster radial gradients, together with compressive features, 
conspire to moderately flatten the perturbations spectrum, slightly departing from the KOC theory and inducing a slow decrease in $A_v/A_\rho$. \\ 

\item
Thermal conduction strongly damps density/entropy perturbations (the spectral steepening occurs where Prandtl $P_{\rm t}<100$), but leaves {\it unaltered} the velocity cascade. This has a dramatic consequence on $A_v/A_\rho$, inverting the downtrend shown in the non-diffusive model.
The ratio can widen up to $\sim$5, as a function of $P_{\rm t}^{-1}\propto f/M$, unveiling the presence of significant conductivity in the ICM, and breaking any degeneracy in the interpretation of single spectra. \\
\end{itemize}

\noindent
The real-space and projected maps carry important informations: 
\begin{itemize}
\item
The ideal or poorly diffusive flows ($f\lta10^{-2}$) show complex filamentary and patchy density/entropy structures (similar in the core and outskirts), excited by the large-scale waves and later altered by hydrodynamical instabilities. The conductive models instead show smooth maps due to the smearing of sharp features, which does not affect the turbulent eddies. 
Albeit sharing similar amplitude, velocities have uncorrelated phases with the tracer, hence a filament does not  necessarily imply a high local velocity. \\

\item
The thermodynamic fluctuations can be described by log-normal distributions, with weak non-Gaussian deviations, strengthening the role of the power spectrum.\\

\item
Synthetic X-ray images of velocity dispersion show that the forthcoming {\it Astro-H} (and {\it Athena}) will be able to well detect subsonic ICM turbulence. Using the broadening of the Fe XXV line, the detectable turbulent broadening will be $\gta 200$ km s$^{-1}$, i.e.~$M_{\rm 1D}\gta0.13$ for massive clusters,
probing density perturbations of the order of a few per cent and allowing to calibrate the linear relation.
The projected velocity maps (line shift) highlight instead the power stored in the large-scale motions, constraining the injection scale.\\

\end{itemize}

The analysis presented in this work shows the wealth of information that can be extracted from the
ICM power spectrum.
For instance, \citet{Schuecker:2004} retrieve in Coma pressure fluctuations which are mildly adiabatic and trace a Kolmogorov spectrum, in line with a $M\sim0.4$ turbulent flow. In \citealt{Gaspari:2013_coma} (see also \citealt{Churazov:2012}), we showed that the density spectrum arising from deep {\it Chandra} data of Coma is consistent with a similar level of turbulence (several 100s km s$^{-1}$), along with highly suppressed conduction ($f\sim10^{-3}$). \citet{Sanders:2012} found density fluctuations ($\lta8$ per cent) having a cascade shallower than Kolmogorov in AWM7 cluster, implying highly suppressed conduction and $M\lta0.18$. 
Being able to quickly convert between thermodynamic properties and gas motions through a simple linear relation, or being able to assess the plasma diffusivity through the spectral slope or the $A_v/A_\rho$ diagnostic,
is a powerful tool for both observational and theoretical study. The same analysis can be extended to other gaseous halos, such as massive galaxies and groups.
Although current constraints are in its embryonic stage, we are beginning to understand the richness of informations that the ICM power spectrum can convey.
Future studies and observations (not only in the X-ray band, but also via SZ maps) will be able to improve and exploit the full potential of the ICM power spectrum, helping us to probe the physics of the gaseous medium with high precision.

\section*{Acknowledgments}
The FLASH code was in part developed by the DOE NNSA-ASC OASCR Flash center at the University of Chicago. 
M.G. is grateful for the financial support provided by the Max Planck Fellowship.
We acknowledge the MPA, RZG, and CLS center for the availability of high-performance computing resources. 
D.N. and E.L. acknowledge support from NSF grant AST-1009811, NASA
ATP grant NNX11AE07G, NASA Chandra grants GO213004B and TM4-15007X,
and the Research Corporation. 
M.G. thanks A. Schekochihin, R. Sunyaev, S. Borgani, F. Brighenti, F. Miniati, D. Eckert, S. Molendi, X. Shi, E. Pointecouteau for helpful comments, and the anonymous referee for a highly positive feedback.

\bibliographystyle{aa}
\bibliography{biblio}

\label{lastpage}

\end{document}